\documentclass[preprint,journal]{vgtc}            

\usepackage[dvipsnames,svgnames]{xcolor}

\usepackage{microtype}                 
\PassOptionsToPackage{warn}{textcomp}  
\usepackage{textcomp}                  
\usepackage{mathptmx}                  
\usepackage{times}                     
\usepackage{cite}                      
\usepackage{tabu}                      
\usepackage{booktabs}  
\usepackage{tabularx}

\usepackage{makecell}
\usepackage{multirow}
\usepackage{colortbl}
\usepackage{array}

\usepackage{lscape}
\usepackage{mdwlist}
\usepackage[,pagebackref,bookmarks]{hyperref}

\usepackage{tikz}
\definecolor{darkred}{HTML}{8b0000}
\usepackage{calc}
\newlength\myheight
\newlength\mydepth
\settototalheight\myheight{Xygp}
\newcolumntype{C}{>{\centering\arraybackslash}m{3cm}}
\newcolumntype{L}{>{\arraybackslash}m{9cm}}

\makeatletter
\def\thickhline{%
  \noalign{\ifnum0=`}\fi\hrule \@height \thickarrayrulewidth \futurelet
   \reserved@a\@xthickhline}
\def\@xthickhline{\ifx\reserved@a\thickhline
               \vskip\doublerulesep
               \vskip-\thickarrayrulewidth
             \fi
      \ifnum0=`{\fi}}
\makeatother

\newlength{\thickarrayrulewidth}
\usepackage{boldline}

\usepackage[most]{tcolorbox}
\usepackage{pifont}
\usepackage{wrapfig}
\usepackage{tabularx}
\usepackage{hhline}
\usepackage{fontawesome}
\usepackage{xspace}

\usepackage[frozencache,cachedir=.]{minted}
\usepackage[english]{babel}
\usepackage{hyperref}   
\addto\extrasenglish{

}

\usepackage{xspace}

\newcommand{\VAE}{\mintinline{yaml}{VAEs}\xspace}
\newcommand{\VAU}{\mintinline{yaml}{VAUs}\xspace}

\usepackage{enumitem}
\usepackage[skip=6pt]{caption}
\usepackage[normalem]{ulem}

\onlineid{0}

\vgtccategory{Research}

\vgtcpapertype{theory/model}

\title{A Heuristic Approach for Dual Expert/End-User\\ Evaluation of Guidance in Visual Analytics}

\author{%
\authororcid{Davide Ceneda}{0000-0003-1198-567X}, \authororcid{Christopher Collins}{0000-0002-4520-7000}, \authororcid{Mennatallah El-Assady}{0000-0001-8526-2613},\\ \authororcid{Silvia Miksch}{0000-0003-4427-5703}, \authororcid{Christian Tominski}{0000-0001-7704-355X} and \authororcid{Alessio Arleo}{0000-0003-2008-3651} }

\authorfooter{
\item
 Davide Ceneda, Alessio Arleo and Silvia Miksch are with TU Wien.  
\item
 Christopher Collins is with Ontario Tech University.
\item
Mennatallah El-Assady is with ETH Zurich, AI Center.
\item
 Christian Tominski is with the VAC Institute, University of Rostock.
}

\abstract{%
Guidance can support users during the exploration and analysis of complex data.
Previous research focused on characterizing the theoretical aspects of guidance in visual analytics and implementing guidance in different scenarios. However, the evaluation of guidance-enhanced visual analytics solutions remains an open research question. We tackle this question by introducing and validating a practical evaluation methodology for guidance in visual analytics. We identify eight quality criteria to be fulfilled and collect expert feedback on their validity. To facilitate actual evaluation studies, we derive two sets of heuristics. The first set targets heuristic evaluations conducted by expert evaluators. The second set facilitates end-user studies where participants actually use a guidance-enhanced system. By following such a dual approach, the different quality criteria of guidance can be examined from two different perspectives, enhancing the overall value of evaluation studies. To test the practical utility of our methodology, we employ it in two studies to gain insight into the quality of two guidance-enhanced visual analytics solutions, one being a work-in-progress research prototype, and the other being a publicly available visualization recommender system. Based on these two evaluations, we derive good practices for conducting evaluations of guidance in visual analytics and identify pitfalls to be avoided during such studies.
}


\teaser{
  \begin{tikzpicture}
        \node[anchor=south west,inner sep=0] at (0,0) {\includegraphics[width=0.99\linewidth]{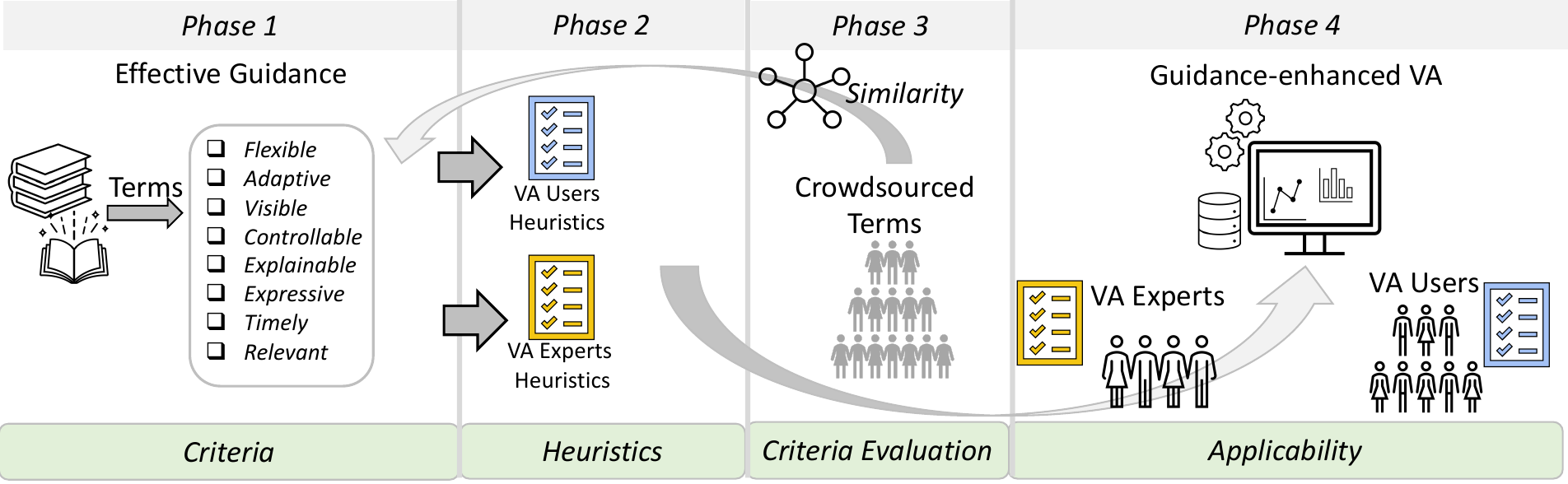}};
        \node[] at (2.5,0.8) {\small \textcolor{darkred}{(Sect. \ref{sec:criteria})}};
        \node[] at (6.3,0.8) {\small \textcolor{darkred}{(Sect. \ref{sec:heuristics})}};
        \node[] at (9.1,0.8) {\small \textcolor{darkred}{(Sect. \ref{sec:eval1})}};
        \node[] at (13.3,0.8) {\small \textcolor{darkred}{(Sect. \ref{sec:eval2})}};

        \node[] at (6.5,1.9) { \href{https://gitlab.cvast.tuwien.ac.at/dceneda/heuristics-guidance-effectiveness/-/blob/main/Designer-Heuristics.pdf}{\includegraphics[height=1em]{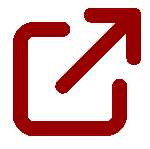}}};
        \node[] at (6.5,3.6) { \href{https://gitlab.cvast.tuwien.ac.at/dceneda/heuristics-guidance-effectiveness/-/blob/main/User-Heuristics.pdf}{\includegraphics[height=1em]{pictures/wraps/hyper-dark-red.png}}};

    \end{tikzpicture}
  \caption{The methodological approach that guided the design of our evaluation methodology for guidance approaches.}
\label{fig:teaser}
}

\graphicspath{{figs/}{figures/}{pictures/}{images/}{./}} 

\usepackage{tabu}                      
\usepackage{booktabs}                  
\usepackage{lipsum}                    
\usepackage{mwe}     

\usepackage{mathptmx}                  

\begin{document}


\firstsection{Introduction}

\maketitle
Visual Analytics (VA) combines human and machine effort to generate insights \cite{keim2008visual, cook2005illuminating}. The problem, though, is that human-computer collaboration is typically imperfect due to noisy and incomplete communication on goals and tasks between user and system, which translates into time-consuming trial-and-error attempts at data analysis. For this reason, guidance approaches are studied to ease the user's analysis and close potential knowledge gaps that might hinder the analysis~\cite{ceneda2017characterizing}.

The literature on guidance approaches is rich and spans multiple fields~\cite{ceneda2019review}. Typical examples of guidance are recommender systems \cite{may2011guiding, luboschikheterogeneity, gotz2009behavior}, which aim to suggest to the user how to analyze the data (e.g., the next action to take), or how to set up the visual environment to facilitate data exploration and completing analysis tasks. In general, guidance approaches can provide a variety of support to resolve knowledge gaps of different types (e.g., for supporting data manipulation \cite{kandel2012profiler}, or to support the classification and modeling of data \cite{choo2010ivisclassifier}) in different situations \cite{kandel2011wrangler, gladisch2013navigation, heer2005vizster}.

Besides practical implementations, guidance has also been studied extensively from a theoretical point of view \cite{ceneda2018decision, collins2018guidance}. Recently, researchers made further steps to bridge this theory to practice, resulting in frameworks and guidelines for designing and implementing effective guidance \cite{sperrle2022lotse, Han23Guidance}.
One of the open issues with guidance is that while existing frameworks and approaches claim that the envisioned guidance should be effective for the user (e.g., \cite{ceneda2020guide}), it is still not very clear how to evaluate in practical terms the extent to which the guidance succeeded in helping the user. While the literature concerning the evaluation of visual interfaces is abundant, the same cannot be said for the evaluation of guidance approaches. 
In other words, evaluating the effectiveness of guidance is an open challenge, with most existing works resorting to (sometimes inappropriate) methodologies initially created for evaluating visualizations, not guidance.

But how can guidance be \emph{effective}? The term is rather vague when referring to guidance, and hence is open to many interpretations. Can we consider it effective when it improves the user's performance? Or when it is provided in a timely manner? 
As it is when evaluating visualizations, 
there are many facets to guidance effectiveness, too.
If we aim to design and implement well-thought guidance approaches, we need to be able to specify with sufficient precision which qualities and characteristics a guidance approach should possess to work properly for the user. Conversely, this could also help practitioners understand why certain approaches fall short of providing effective support, which could lead to the design of better guidance in a broader sense.

To address these challenges, we propose a heuristic approach to evaluating guidance in VA. Building on a recent attempt at characterizing effective guidance design \cite{ceneda2020guide}, we conducted extensive literature research and interviewed visualization experts to identify what characteristics make guidance effective, ultimately resulting in eight quality criteria. These were then instantiated into two sets of heuristics and questions that can be answered by \textit{VA experts} (\raisebox{-0.8\mydepth}{\includegraphics[height=1em]{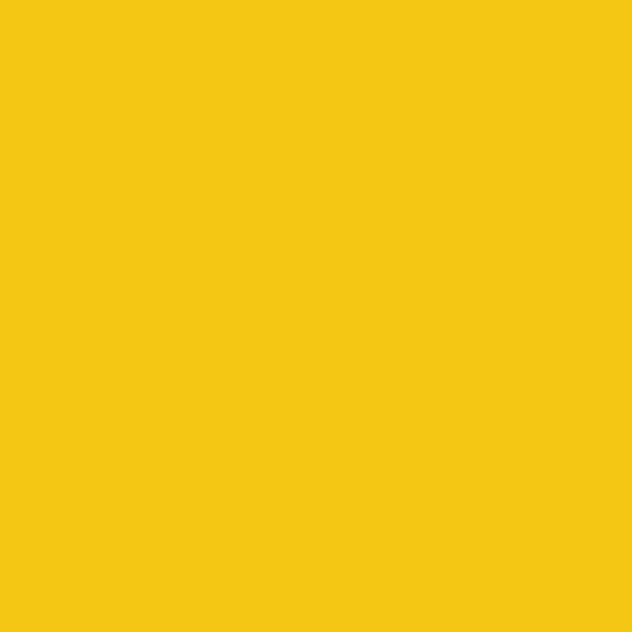}} \VAE{}), such as designers, to assess the quality of the design, and by \textit{VA users} (\raisebox{-0.8\mydepth}{\includegraphics[height=1em]{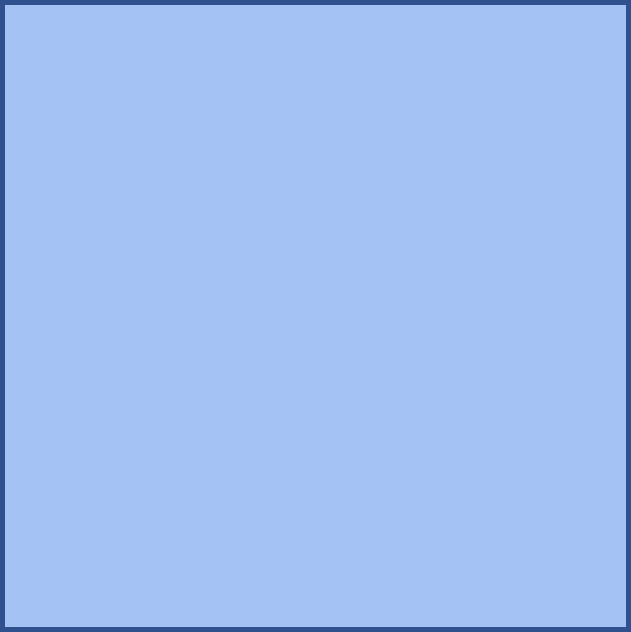}} \VAU{}), to assess if the guidance was effective for them during the analysis when solving tasks -- hence the ``dual'' nature of our methodology.

By evaluating guidance from the perspective of \VAU{} and \VAE{}, we aim to provide a comprehensive quality assessment of the guidance.
Summarizing, the paper's contributions are as follows: 

\vspace{-2mm}
\begin{itemize}[leftmargin=*]\setlength\itemsep{-0.1em}
    \item A characterization of the \textit{effectiveness} of guidance through the identification of a set of eight quality criteria (see Tab.~\ref{table:heuristics}, left column) that contribute to the value and to the effectiveness of guidance. 
    \item Two sets of heuristics (see Tab.~\ref{table:heuristics}, center and right column) to evaluate and quantify the effectiveness of guidance, either from the perspective of VA experts (e.g., designers) or VA users. 
    \item An in-depth evaluation of our methodology using two guidance-enhanced VA systems, and a qualitative analysis of the results.  
    \item A detailed protocol and templates \href{https://gitlab.cvast.tuwien.ac.at/dceneda/heuristics-guidance-effectiveness}{(download them here \raisebox{-0.8\mydepth}{\includegraphics[height=1em]{pictures/wraps/hyper-dark-red.png}})} to replicate the study and apply it to other guidance approaches.
\end{itemize}

\section{Related Work}
\label{sec:related}
In this section, we review the literature on the evaluation of visualizations and guidance.

\subsection{Evaluating the Effectiveness of Visualizations} Studying the factors that contribute to the effectiveness of visualizations is a challenge tied to the development of evaluation methodologies. 
By evaluating them systematically, we can judge visual interfaces and get a glimpse of their potential and limitations. In other words, evaluation methodologies aim to assess the effectiveness of a given information display in supporting specific perceptual tasks \cite{mackinlay1986automating}.

The effectiveness of visual displays is the result of many different aspects concerning how the information is shown to the user. 
Miksch and Aigner argue that the \textit{``effectiveness primarily considers not only the degree to which a visualization addresses the cognitive capabilities of the human visual system, but also the task at hand, the application background, and other context-related information, to obtain intuitively recognizable and interpretable visual representations''} \cite[p.287]{miksch2014matter}.
These statements highlight why evaluating visualizations is a challenging task, and that accounts for the variety of domains and scenarios in which information visualization tools are used \cite{shneiderman2006strategies}.

Among the many strategies to evaluate visualization, one of the most common is to measure the performance of multiple users using one or more tools in controlled experiments~\cite{plaisant2004challenge}. With this methodology, evaluators observe and collect metrics on how users perform data analyses and use a given tool. Typically, a baseline is used as a benchmark against which all the results are compared. Multiple different metrics can be utilized to assess the usability and performance of visual tools~\cite{chen2000empirical,chen2000empirical1}. For instance, execution time or error rates can be utilized to assess a visualization's effectiveness, i.e., the accuracy fostered by the visualization. Beyond accuracy, assessing effectiveness entails measuring the memorability of a visual display, i.e., recalling data cases or elements of the interface, as well as collecting qualitative and subjective aspects of visualization tools such as user's enjoyment, satisfaction, fun~\cite{saket2016beyond}, and aesthetic pleasure~\cite{he2022beauvis}.


An additional way for evaluating visualizations is using heuristics to assess usability and identify subjective and qualitative problems that might hinder the completion of the analysis~\cite{zuk2006heuristics}. The use of heuristics is common in human-computer interaction to test whether tools comply with well-known design principles and rules of thumb~\cite{forsell2012evaluation}. During the years, many heuristics have been identified to evaluate visualizations in specific contexts and domains~\cite{scapin1997ergonomic, nielsen2005ten, wang2000guidelines, zuk2006theoretical, amar2004best, forsell2010heuristic}. Molich and Nielsen identified nine principles and used them to evaluate visual interfaces, organizing usability issues according to their severity~\cite{molich1990improving}. Heer et al.~\cite{heer2007animated} identified a set of heuristics to evaluate the effectiveness of animated transitions between well-known chart types. Mankoff et al. describe a set of heuristics for the evaluation of ambient displays, which are aesthetic visualizations portraying non-critical information on the periphery of a user's attention~\cite{mankoff2003heuristic}. Wall et al. describe a set of heuristics to evaluate and quantify the value of visualizations, assessing their ability to communicate the \textit{essence} of the data, the quality of the \textit{insights}, the \textit{confidence} of the user, and the performance and \textit{time} of the analysis (i.e., the so-called ICE-T methodology)~\cite{wall2018heuristic}. An interesting aspect of this methodology is that it provides a solid way to compare the qualities of multiple (even different) visualization approaches.

In summary, the effectiveness of visualization tools is the result of a variety of factors that mostly depend on how the visualization approach was designed and also on how people use and experience the tool. Taking inspiration from evaluation methodologies for visualizations, we aimed to identify what factors contribute to the effectiveness of guidance. In this aspect, our approach is similar to the ICE-T approach by Wall et al.~\cite{wall2018heuristic} in that we aim to provide a clear numerical value to represent the effectiveness of guidance approaches and enable the comparison of their characteristics. Moreover, similar to the work by Stasko\cite{stasko2014value}, we aim to characterize but also quantify the value of guidance in VA and the multiple facets of guidance effectiveness.

\subsection{Evaluating the Effectiveness of Guidance} In parallel with researching visualizations, scientists also looked for approaches to support and enhance their use. In this regard, it makes sense to differentiate between \emph{onboarding} approaches~\cite{stoiber2022perspectives} -- which are mostly utilized before the analysis (e.g., videos and tutorials) to facilitate their use and \emph{guidance} approaches~\cite{ceneda2017characterizing} --  which are typically used during the analysis to enhance and facilitate the completion of tasks. In this paper, we focus on the evaluation of guidance techniques.

A guidance approach can be characterized by the knowledge gap(s) it aims to address, by the input used and output produced, and by the degree, which represents the amount of support given to the user~\cite{ceneda2017characterizing}. Collins et al. expand the discussion of guidance and examine additional types of knowledge gaps~\cite{collins2018guidance}. Ceneda et al. review the literature concerning guidance approaches categorizing papers according to the guidance provided by the system to the user and according to the guidance the system receives from the user~\cite{ceneda2019review}. Sperrle et al. expand this categorization, focusing on learning and teaching processes~\cite{sperrle2020learning}. Han and Schulz investigate a methodology for implementing guidance based on decision support theory~\cite{Han23Guidance}.
 
When analyzing the literature on guidance approaches, we can see no single methodology for their evaluation. Even more, the evaluation of guidance and visualization tools are typically carried out together. Hence, the effects and the contribution of guidance to the overall analysis process are not immediately clear or recognizable. Heer et al. describe an approach to enhance the exploration of communities in large networks~\cite{heer2005vizster}. To evaluate the approach, the authors focus on its usability~\cite{nielsen2005ten}, following a standard methodology to evaluate visualizations. When evaluating guidance, most approaches focus on assessing whether the introduction of guidance leads to performance improvements. For instance, Bouali et al. present a tool that can recommend appropriate visual encodings to the user; the approach is evaluated in a controlled experiment comparing the performance of users receiving guidance with those applying a completely manual workflow~\cite{bouali2016vizassist}. Gotz et al. also consider performance improvements (i.e., (task) completion time and error rate) to evaluate their guidance-enhanced approach to suggest appropriate visualizations based on user interaction trails~\cite{gotz2010harvest}. May et al. describe an approach to guide the exploration of large graphs using signposts. To evaluate it, the authors utilize completion time and two click-based metrics, showing how guidance reduces them significantly. Wongsuphasawat et al. focus on the effects of guidance and report how the use of a recommender system encourages the use of visualizations of interest~\cite{wongsuphasawat2017voyager}.
 
Beyond raw performance, Ceneda et al. explore the effects of guidance on users, describing how, in particular, the guidance provided changed their exploration strategies, i.e., the way users analyzed the data and how, conversely, this led them to a deeper understanding of the data and increased confidence towards the results obtained~\cite{ceneda2018guided}. Krause et al. assess how their guidance-enhanced approach can foster deeper insights into predictive models and improve model interpretability~\cite{krause2016interacting}. An interesting way to evaluate guidance is presented by Krishnamoorthy and Brusilovsky, who presented a tool to guide students toward appropriate examples of a pre-selected topic~\cite{krishnamoorthy2006personalized}. In particular, the authors focus not only on the evaluation of the interface design, but also on measuring the amount of help provided to the users, their ability to perceive progress, the navigability of the tool, and on measuring whether a certain goal was achieved thanks to the guidance. O'Donovan et al. propose a qualitative evaluation of their guidance-enhanced approach to improve visualization design with layout suggestions and highlight how the designs recommended by the guidance system were favorably evaluated by experts \cite{o2015designscape}.

We emphasize that similar to what happens for visualizations, guidance approaches are applied in a variety of scenarios providing diverse  types of support. Consequently, no single way to evaluate guidance is known in the literature.
Some approaches use mere performance as an indicator of the effectiveness of guidance. Others check whether an approach complies with suitable design practices, such as usability, recall, and interpretability of the provided support, assessing guidance from a designer's perspective. Finally, others are instead concerned with evaluating the qualitative aspects of guidance and its influence on the analysis, evaluating the provided guidance from a user perspective.

Drawing from the literature, our goal is to derive an evaluation methodology that could shed light on the effectiveness of guidance, taking into account these multiple perspectives and providing a comprehensive assessment and a common ground for the comparison of multiple types of guidance. For this reason, we go a step further than existing approaches, describing two sets of heuristics to evaluate guidance from the perspective of \VAE{} and \VAU{}. The first step in the process to derive our methodology was to identify the factors (i.e., a set of quality criteria) that contribute to making guidance effective. In the following, we describe this process and its outcome.

\section{Methodology}\label{sec:methodology} We followed an iterative process (see Fig.~\ref{fig:teaser}), inspired by He et al.~\cite{he2022beauvis}, to derive our heuristic approach in four phases:
(i) Identification of quality \textbf{Criteria}  (see Sect.~\ref{sec:criteria}).
 (ii) Instantiation of the  \textbf{Heuristics} (see Sect.~\ref{sec:heuristics}).
(iii) \textbf{Criteria Evaluation} (see Sect.~\ref{sec:eval1}).
    (iv) \textbf{Applicability} of the heuristics (see Sect.~\ref{sec:eval2}).


We began our research by conducting a literature review to identify what criteria were used in the literature to evaluate visualizations and guidance approaches. Our goal was to pinpoint the terms scientists associated most often with the effectiveness of an approach. In this initial phase, we collected a comprehensive set of terms, including terms like trustworthy, efficient, appropriate, and accessible. Afterward, we organized the terms in a similarity graph, where terms were represented as nodes and their similarity was shown by edges, to identify overlaps among similar terms relating to potential effectiveness criteria and evaluation methodologies. We used the graph to filter down the initial set of quality criteria and organized them into categories, structuring them around the requirements for guidance design by Ceneda et al.~\cite{ceneda2020guide}. In the following phase, we reviewed the criteria and refined the structure of items and terms in a collaborative manner. This process led to the identification of eight quality criteria that contribute to the effectiveness of guidance. We instantiated these criteria in two sets of heuristics for use by \VAU{} and \VAE{} to evaluate guidance.

We conducted three evaluations of our heuristic approach. The first evaluation was focused on evaluating the criteria. We sent out a questionnaire to more than 80 VA scholars with experience in working with guidance and designing visualizations. We received answers from 25 of them, whose feedback helped us refine the criteria and their characterization and provided us with evidence of the soundness of our methodology. The focus of the second and third evaluations was on the applicability of our approach in practical scenarios. We first asked a group of \VAE{} to apply the heuristics to evaluate existing guidance approaches. Second, a group of \VAU{} worked with a guidance-enhanced tool and subsequently filled out our heuristics-based questionnaire to evaluate the effectiveness of the guidance they received.


\section{The Effectiveness of Guidance} The first step in building an evaluation approach for guidance was identifying the factors and the criteria contributing to guidance effectiveness on which to ground the methodology. A preliminary set of criteria was gathered by reviewing the literature.

\paragraph{Literature review} We searched the proceedings of major conferences as \textit{Conference on Human Factors in Computing Systems (CHI)}, \textit{IEEE VIS} and \textit{EuroVis}, and journals like \textit{TVCG}, \textit{CGF}, using terms like ``guidance," ``visualization," ``evaluation," as well as synonyms and related words. We identified more than a hundred papers dealing with visualizations, evaluation methodologies, and heuristics and a slightly smaller number of papers dealing with guidance. 
In particular, we considered literature reports and surveys (see Sect.~\ref{sec:related}) from which we extracted terms and criteria and categorized papers according to whether they discussed ``metrics" or ``heuristics". For guidance-enhanced approaches, we started our research from the survey by Ceneda et al.~\cite{ceneda2019review}, which we expanded to consider more recent works. Since we could not ground our research on existing evaluation methodologies for guidance, we instead extracted terms and derived qualitative criteria directly from the text, also analyzing the reported user feedback. Typically, guidance approaches are evaluated using qualitative measures. Studies often include direct feedback and citations of the participants (e.g., ``\textit{thanks to the guidance, the participant reported an increased trust in the results obtained}" \cite[p.7]{ceneda2018guided}). In these cases, we synthesized participants' feedback, organizing it in tokens (e.g., ``increased trust'') representing different facets of guidance effectiveness. From this analysis, we collected more than 180 terms comprising heuristics, metrics, and quality criteria, of which around 40 were specifically used for evaluating guidance-enhanced approaches. We did not immediately discard qualities used solely to characterize the effectiveness of visualization approaches. Instead, we kept them to understand whether they could also be applied to guidance, given the tight relationship between the two.

\begin{figure}[ht]
    \centering
    \includegraphics[width=0.99\linewidth]{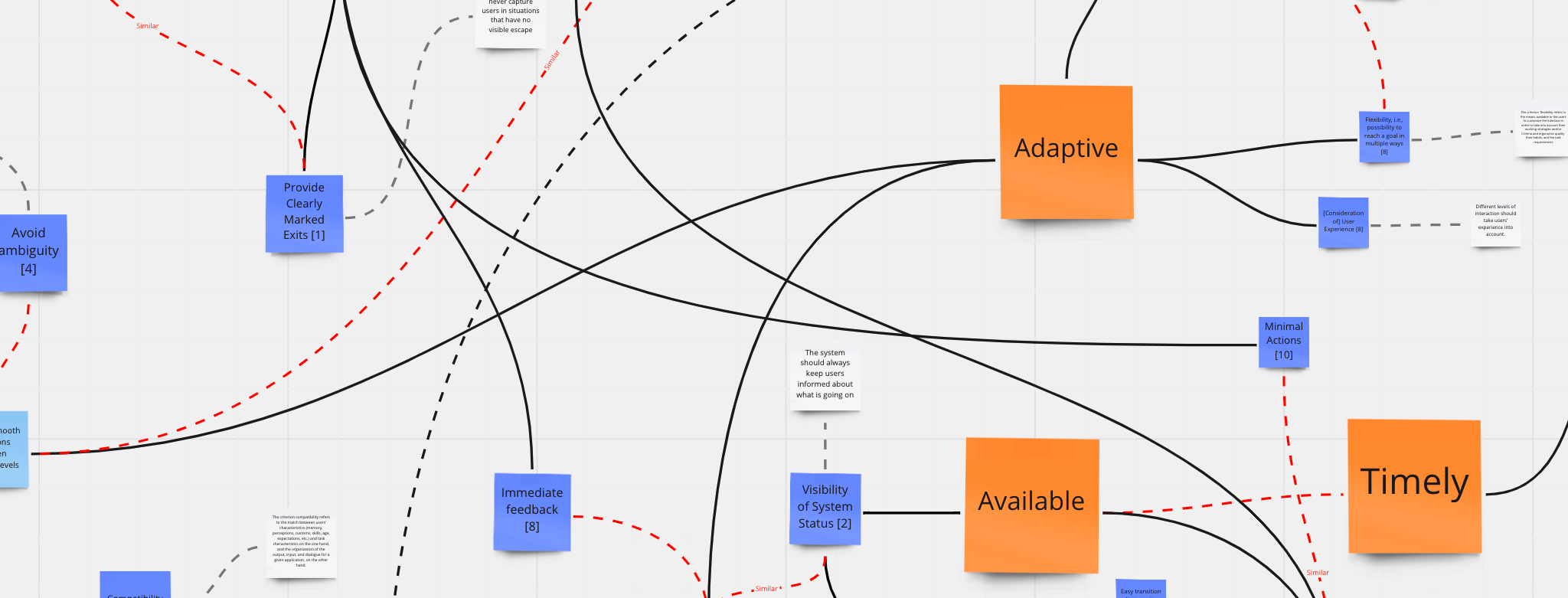}
    \caption{The similarity graph of terms, used to derive a minimal set of quality criteria. Nodes represent qualities and edges represent the similarity between concepts. The full graph is provided as supplementary material.} 
    \label{fig:miro-graph}
\end{figure}

\paragraph{Organizing and structuring} Next, we worked to give structure to the terms collected and organized them in a graph. This graph revealed similarities that allowed us to reduce the number of terms to a manageable level. In 2020, Ceneda et al.~\cite{ceneda2020guide} presented a framework to support the design of guidance approaches comprising steps, questions, and threats that designers must address. Their framework is supported by a list of requirements representing ideal qualities that a guidance approach should have in order to effectively address the users' knowledge gaps. According to this previous work, effective guidance derives from the following qualities: \emph{Adaptive}, \emph{Available}, \emph{Trustworthy}, \emph{Controllable}, and \emph{Non-disruptive}~\cite{ceneda2020guide}. We used these design requirements as initial aggregation nodes for the terms we collected reviewing the literature. We used the online visual collaboration platform Miro to create a similarity graph (with 70+ nodes and 90+ edges). Fig.~\ref{fig:miro-graph} shows a portion of the graph (the complete similarity graph is available as supplementary material). Nodes (in blue) represent the terms and the qualities we collected, which we organized around the requirements described by Ceneda et al. (in orange). Edges between nodes represent possible analogies between concepts. We used this graph to refine and filter the list of terms, equally in a top-down and bottom-up manner, but also to capture the different nuances of guidance effectiveness. For example, thanks to this structuring, we realized that the timeliness of the guidance and the expressiveness of its encoding contribute to its non-disruptiveness. Similarly, we found that the trustworthiness of guidance depends on its explainability and, therefore, merged the two qualities. The result of this phase was a set of 15 terms and qualities, which we further examined in the following phase.

\paragraph{Refining the criteria} \textcolor{black}{While the previous phase was run mainly by two authors, the whole team collaboratively reviewed and examined further the resulting set of terms and qualities. In particular, additional similarities among terms were identified and some of them (e.g., non-disruptiveness) were deemed misleading, and thus, deleted. Hence, the terms were additionally aggregated, and the different nuances related to each quality were captured and formalized in a set of eight quality criteria capturing the effectiveness of guidance.} Each criterion was, in the end, represented (or summarized) by a single term. To avoid confusion and potential ambiguities, we also provide a definition of each term.

\subsection{Quality Criteria and Definitions}
\label{sec:criteria}
Here, we summarize our definitions and characterize the qualities that best describe the effectiveness of guidance. Note that what follows already includes the changes derived from the expert feedback gathered during the evaluation of our methodology (see Sect.~\ref{sec:eval1}).


After listing their definitions, we characterize each quality in more detail and refer to relevant literature in separate paragraphs.

\begin{itemize}[leftmargin=*, itemsep=-0.2em]
    \item \textbf{Flexible} -- The guidance is flexible when the degree of support adequately adapts to the analysis situation(s).
    \item \textbf{Adaptive} -- The guidance is adaptive when it considers user preferences, habits, and current task requirements.
    \item \textbf{Visible} -- The guidance is visible when its recommendations and status can be clearly distinguished in the visual environment.
    \item \textbf{Controllable} -- The guidance is controllable when the user can switch the guidance on/off, request alternative recommendations, or revert previously followed recommendations.
    \item \textbf{Explainable} -- The guidance is explainable when its recommendations are easily understood and the way they were generated is made transparent to the user (also on demand).
    \item \textbf{Expressive} -- The guidance is expressive when its encoding is appropriate and users can extract the information needed to make analytic progress.

    \item \textbf{Timely} -- The guidance is timely when it is provided on time and only when needed.

    \item \textbf{Relevant} -- The guidance is relevant when it guides users toward their analytical goals and supports the completion of tasks.
\end{itemize}
\vspace{-2mm}


\paragraph{Flexible} The extent and type of support the user receives during the analysis (i.e., the \emph{guidance degree}~\cite{ceneda2017characterizing}) are dynamically adjusted. This criterion explicitly addresses the need to identify a suitable {guidance degree and the extent to which the system can adjust the amount and the type of support during the analysis (i.e., how tailored and stringent the guidance within a user task). Typically, the amount of guidance the user needs varies during the analysis, so the system must react to it dynamically to support the user effectively.

\paragraph{Adaptive} \textcolor{black}{The content of the guidance (i.e., the information provided to the users to guide them) is adapted according to the analysis context, for example, when the user task or preferences change\cite{sperrle2022lotse}. While the previously-described flexibility criterion addresses the need of identifying the appropriate \textit{degree} of guidance, this one addresses the need to adapt the \textit{content} of the suggestions given to the user. Typically, adaptiveness can be achieved by considering analytic provenance information, user preferences and user experiences, habits, and interactions with previous suggestions (e.g., rejected guidance) \cite{scapin1997ergonomic}, which can all be used to generate user-tailored content.}

\paragraph{Visible} This criterion ensures that the status of the guidance (e.g., if it is active or not, and its parameters) is visible to the user using appropriate visual feedback and within reasonable time~\cite{scapin1997ergonomic}. That is, the guidance is effective if the user can perceive it. Similar criteria are typically used also for visualizations (e.g., \cite{nielsen2005ten}). In this, we extend their application to guidance-enhanced approaches as well.

\paragraph{Controllable} It must be possible for the user to fine-tune and control the parameters that influence the way the guidance is produced~\cite{ceneda2020guide}. This criterion aims to evaluate whether a guidance approach allows the user to control, for instance, the level of details and the amount of guidance provided (see \textit{flexibile}), and switch it off occasionally. Drawing from well-known heuristic approaches, an effective guidance system should show marked exits, support undo/redo of actions \cite{molich1990improving}, allow easy access to parameters and settings \cite{nielsen2005ten} and, more in general, to orientation features, for improving usability and hence, guidance effectiveness.

\paragraph{Expressive} This criterion aims to assess whether guidance suggestions can be provided and communicated to the user using an appropriate language (i.e., using the user's language \cite{molich1990improving}). The chosen encoding should be used consistently throughout the system so to avoid confusion and ambiguities and to make the guidance recognizable in the visual interface (compare, for instance, the \emph{legibility} heuristic~\cite{scapin1997ergonomic}). The expressiveness of guidance also concurs with its explainability.

\paragraph{Explainable} This criterion aims to assess the way the guidance is communicated and understood by the user. Explainable suggestions concur to increased acceptance and trust (see guidance trustworthiness~\cite{ceneda2020guide}) towards the received guidance. Furthermore, to uphold effectiveness, the explainability should not be limited to the suggestion itself but also clearly expose the process that led the system to it (which is typical in ML applications~\cite{spinner2019explainer}). The user should be entitled to ask for more explanations and details, if necessary (see the \emph{controllable} criterion). Generally, the guidance should adhere to the principle of least astonishment\cite{BERGERON1972175}, i.e., the system should behave in a way that most users will expect. When this does not happen (e.g., when the user has to be addressed toward a new analysis path), each action and suggestion coming from the system should be adequately explained to avoid confusion.

\paragraph{Timely} \textcolor{black}{The timeliness criterion assures that the guidance is provided at the right time, without delays, so as not to disturb or interrupt the user, or interfere with other tasks (e.g., when the task changes and the guidance is late and, hence, not relevant anymore). Together with the use of appropriate explanations, the timeliness of guidance concurs to make it accepted and trusted by the user \cite{ceneda2021show}}.

\paragraph{Relevant} The last criterion deals with assessing whether the guidance can guide the user toward relevant analysis results and, at the same time, avoid that the user makes errors in the process. The guidance should have an aim and be able to perform basic tasks to support the user. The relevance of guidance is related to its ability to achieve these aims and perform these tasks. Typically, the system should be able either \textit{to guide} users toward the analytical goal, to \textit{correct} and readdress them along other analytical paths if problems are detected, and \textit{to ask} for user input if needed (see the \emph{guide}, \emph{correct}, and \emph{ask} tasks of P\'{e}rez-Messina et al.~\cite{perez2022typology}). The analysis process is typically complex, and due to this complexity, many mistakes can happen. Drawing from accuracy metrics, the system should help users avoid analysis errors, or at least help them recover and get back on track, exposing uncertainties and biases ~\cite{amar2004best}. In this regard, the combination of error detection and prevention functionalities accompanied by appropriate explanations (see \textit{explainable} and \textit{expressive} criteria) is necessary to ensure the user does not return to the erroneous analysis path. 
The guidance should also lead users towards their analysis goals, help them discover the unexpected when possible, or even generate hypotheses about the data in the first place. If a goal cannot be immediately identified (e.g., at the beginning of the analysis, goals could be undefined), the system can poke the user for details about the scope and goals (see the ``ask'' tasks~\cite{perez2022typology,sperrle2022lotse}). All these results could be achieved in multiple ways. Hence, the system should support the exploration of alternatives when several actions are possible depending on the analysis context. The ability of the system to achieve these guidance tasks contributes to its relevance.

\begin{table*}
    \centering
    \advance\leftskip-0.5cm
    \includegraphics[width=1.08\textwidth, height=\textheight]{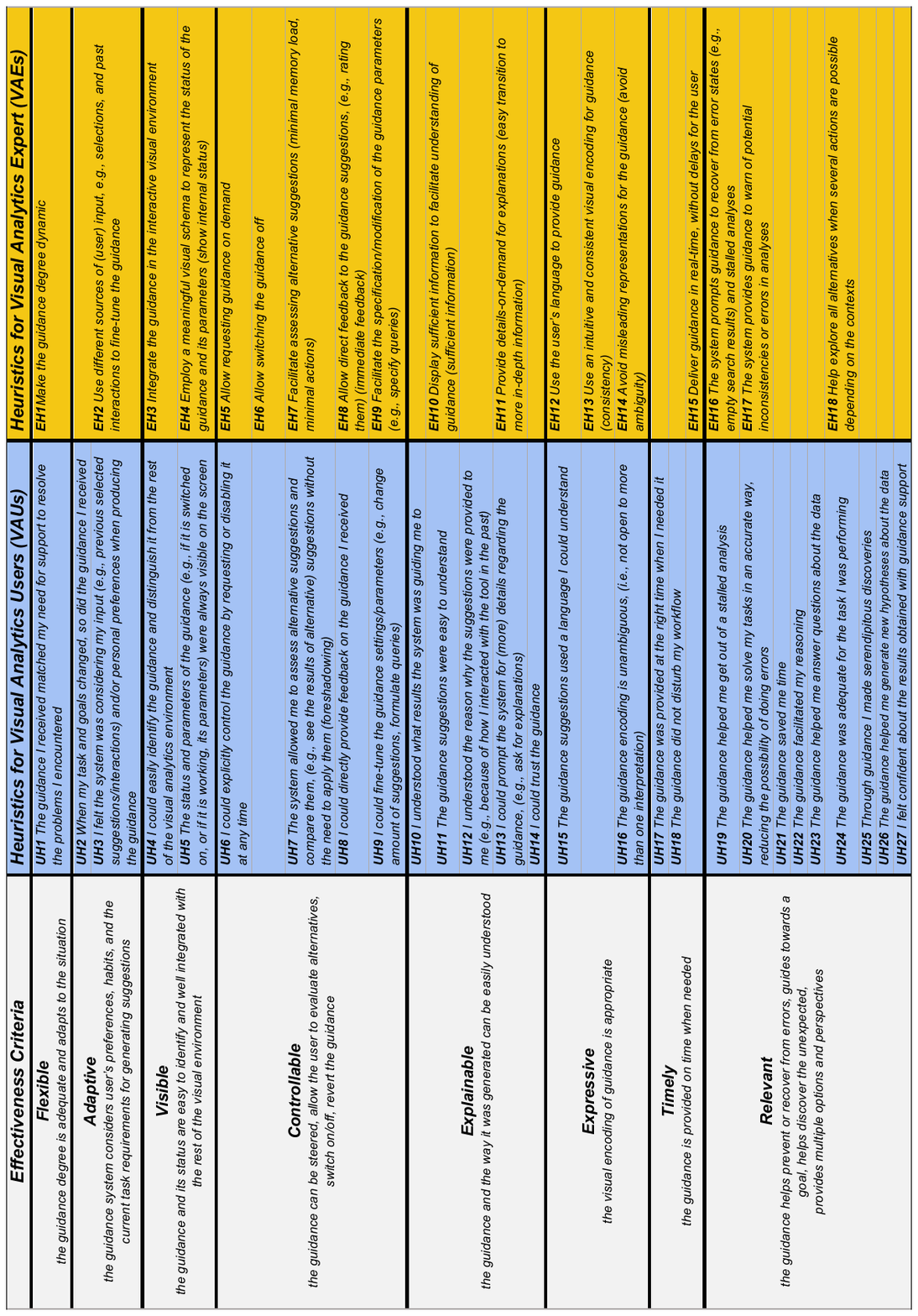}
    \caption{The sets of heuristics to evaluate the effectiveness of guidance. The columns list the quality criteria associated with the effectiveness of guidance (left), the \VAU{} heuristics (center), and the \VAE{} heuristics (right). Rows specify the actual heuristics.}
    \label{table:heuristics}
\end{table*}

\subsection{Heuristics for Dual Expert/End-User Evaluation}\label{sec:heuristics} \textcolor{black}{We derived the heuristics from our quality criteria such that the resulting heuristics covered all the nuances specified by the criterias' definitions.} The heuristics are listed in the table in Tab.~\ref{table:heuristics} in the  \raisebox{-0.8\mydepth}{\includegraphics[height=1em]{pictures/wraps/blue_icon.png}} central and \raisebox{-0.8\mydepth}{\includegraphics[height=1em]{pictures/wraps/yellow_icon.png}} right columns along with the quality criteria in the left column.

\paragraph{Evaluators} To whom is our methodology addressed? During the initial phases of research, we realized that certain aspects of guidance could be better judged after having experienced the tool and received the guidance in the first place. For example, to check if the guidance was provided on time or helped in getting out of problematic situations, it makes sense to refer to VA users who actually received the guidance. Likewise, we realized that VA experts (e.g., designers) could better assess other criteria from the first stages of development without needing to actually experience the tool. This led us to derive and instantiate the criteria in not one but two sets of heuristics, each focused and tailored for a specific type of user: \VAE{} and \VAU{}.

In total, we describe a different number of heuristics, 27 to be assessed by \VAU{} and 18 for \VAE{}. We expect the expert evaluators to be knowledgeable of the guidance concept and of visualization design in general. We do not expect the user evaluators to be knowledgeable besides being able to use a VA system.

\paragraph{Formulation} We formulated the heuristics keeping in mind their specific audience. For instance, this implied formulating the \VAU{} heuristics using the first-person singular and avoiding domain-specific terms and references to guidance terminology (e.g., references to the guidance degree or the knowledge gap were avoided), as we do not expect \VAU{} to possess any specific knowledge of those terms. Also, \VAU{} heuristics were formulated positively using single sentences and avoiding conjunctions, to ease their interpretation and their assessment in a survey. Conversely, heuristics meant for \VAE{} are shorter and typically formulated as rules of action.

\subsubsection{Running the evaluation}\label{sec:heuristics-use} Our heuristics are to be applied practically using different questionnaires for \VAE{} and \VAU{}. While \VAU{} need to have direct access to the studied system and its guidance to evaluate the effectiveness, \VAE{} do not necessarily need to use the tested tool (i.e., the system could still be in the design phase) but may rely instead on design descriptions (e.g., sketches, videos, mockups). 

Given the different target audiences of the two heuristics sets, we expect them to be rated differently according to the following scheme.

\paragraph{VA experts} \VAE{} can rate the system's adherence to the heuristics assessing the application or the violations of the different design principles (e.g., using a Likert scale from 1--clearly violated to 7--clearly applied, or N/A not applicable). The identified violations of design principles and heuristics can be reported. Subsequently, severity ratings \cite{nielsen2005ten} (e.g., from 0 -- not a problem, to 4 -- design catastrophe) could be used to assess their severity and allocate enough resources to possibly solve them to improve guidance effectiveness. In line with Molich and Nielsen, we expect \VAE{} heuristics to be rated at least by 3-5 evaluators to identify a wide range of design problems. Finally, given the way we created the heuristics, we assume that the evaluation of guidance can be run in parallel with the typical assessment of usability problems\cite{molich1990improving}.

\paragraph{VA users} On the other hand, as \VAU{} heuristics are formulated as statements, we expect \VAU{} to rate them using a 7-point Likert scale ranging from 1-strongly disagree to 7-strongly agree or N/A not applicable, to assess their agreement with the reported statement/heuristics. Subscores can be calculated for each criterion by summing the agreement (1-7), representing an estimation of the different qualities of the system under analysis. A final score can be obtained by summing the subscores to obtain an overall estimation of the effectiveness of guidance. Calculating and using scores and subscores enables the direct comparison of (different) guidance-enhanced approaches and their characteristics.
Finally, \VAU{} heuristics should be evaluated with at least 15 users in order to achieve a reliable result. 

\paragraph{Questionnaires} The questionnaires for \VAE{} and \VAU{}, and the protocol to run the evaluations can be found \href{https://gitlab.cvast.tuwien.ac.at/dceneda/heuristics-guidance-effectiveness}{in this Git repository \raisebox{-0.8\mydepth}{\includegraphics[height=1em]{pictures/wraps/hyper-dark-red.png}}} \cite{gitgit}.

\section{Consolidating Quality Criteria}
\label{sec:eval1}

As part of our effort to assess the quality, clarity, and comprehensiveness of our set of effective guidance criteria,
we ran an expert evaluation with a twofold purpose.
First, we wanted to gather feedback about the selection of terms, on the lookout for what we might have missed during the literature review (see Sect.~\ref{sec:methodology}).  
Second, we were interested in the participants' opinions about the quality criteria and their definitions, looking for alternate formulations and/or additional quality criteria.
Here, we describe the study design that led us to those results.

\paragraph{Procedure} The evaluation was arranged as a two-part online survey. After completing the informed consent form, we described the scope of the study to the participants. In the first part of the study, we asked the participants to think of at least three adjectives or qualities they associated with their personal idea of effective guidance. We also asked, as an optional question, to provide and describe additional qualities, discuss examples, or further justify the given adjectives.
In the second part, we presented visualization experts with nine quality criteria with their proposed definitions (different from the final list presented in Sect.~\ref{sec:criteria} and available in the supplemental material). For each one of them, we first asked the participants their opinion about their relevance in the context of a VA system on a 1 (not important at all) to 5 (essential) scale, e.g., ``\textit{In the context of a VA system, in your opinion how relevant is it for the provided guidance to be adaptive?}''. The participants were also asked to evaluate our definitions of the criteria, stating if they agreed with them or not. Participants were encouraged to provide their own definitions and additional comments about their assessment. Finally, participants could provide overall feedback on the set of criteria as a whole, evaluating, for instance, the importance of each criterion compared to the others in capturing the effectiveness of guidance.

\paragraph{Participants} We administered an online questionnaire to scholars with proven research experience (i.e., a publication record) in visualization and guidance. We contacted them by email and gave them three weeks to complete the survey. Participants were not remunerated for their participation. Ultimately, 25 of them answered the questionnaire.

\paragraph{Outcome} From the answers, we gathered around 100 terms, qualities, and adjectives that experts associated with effective guidance. We preprocessed the list to remove duplicates and merge similar terms. This step was facilitated by the examples and explanations provided by the participants. Overall, we obtained around 50 terms to merge with those collected from our initial literature review (see Sect.~\ref{sec:methodology}). For this purpose, we created a similarity graph in which the crowdsourced qualities (nodes) were associated (with edges) with previously gathered literature terms according to their similarity. 

The similarity graph helped us refine the criteria and capture nuances we did not previously consider.
For example, participants reported confusion when reading the definitions of \emph{adaptive} and \emph{flexible} guidance and suggested exchanging their names. Other fine-grained adjustments were made to the other criteria and their definitions. For example, based on the feedback received, we renamed a criterion to \emph{relevant} (which previously was referred to as \textit{purposeful}) and merged it with one concerning the \textit{resiliency} of guidance. 
Finally, it is worth mentioning that the study participants mostly associated guidance effectiveness with its \textit{explainability} (i.e., among the reported terms, many cited ``clear,'' ``transparent,'' "descriptive,'') and its \textit{relevance} (i.e., ``successful,'' ``insights,'' ``task-oriented,'' ``helpful,'' ``meaningful''). This information can, for example, be helpful for weighting the severity ratings in actual evaluations of guidance systems.

All the feedback and insights we gathered from this first study went into the refinement of our quality criteria and respective definitions (see Sect.~\ref{sec:criteria}).

\section{Evaluating Heuristics in Practice}
\label{sec:eval2}
Having formalized criteria after our first expert evaluation, 
we aim to evaluate how our heuristics perform in practice 
in assessing the guidance quality of existing VA systems or in providing insights that can contribute to their design. For this purpose, we designed and ran two more studies, one involving 5 \VAE{}~and another with 39 \VAU{}, described in the following.

\subsection{\raisebox{-0.8\mydepth}{\includegraphics[height=1em]{pictures/wraps/yellow_icon.png}} \VAE~Evaluation} 

The goal of this expert evaluation was to determine whether \VAE{} could apply our heuristics to assess the guidance quality
 in the design stage. 

\subsubsection{Study design}
According to Nielsen \cite{nielsen2005ten}, it is often the case that evaluators do not have access to the actual system when evaluating the ongoing design, especially if the evaluation takes place early in the development cycle. 
Hence, to recreate the conditions of a realistic evaluation of a guidance-enhanced VA prototype, we assembled a video of a system under development, which we then showed to the study participants.

The video showed six scenarios of how later users are supposed to use and benefit from the guidance features included in the system. The system in question is currently in development as part of another research project involving some co-authors of this paper. It can provide different types of guidance at different stages of the data exploration process, and is portrayed in Fig.~\ref{fig:wizard-tool}.

\begin{figure}[t]
    \centering
    \includegraphics[width=\linewidth]{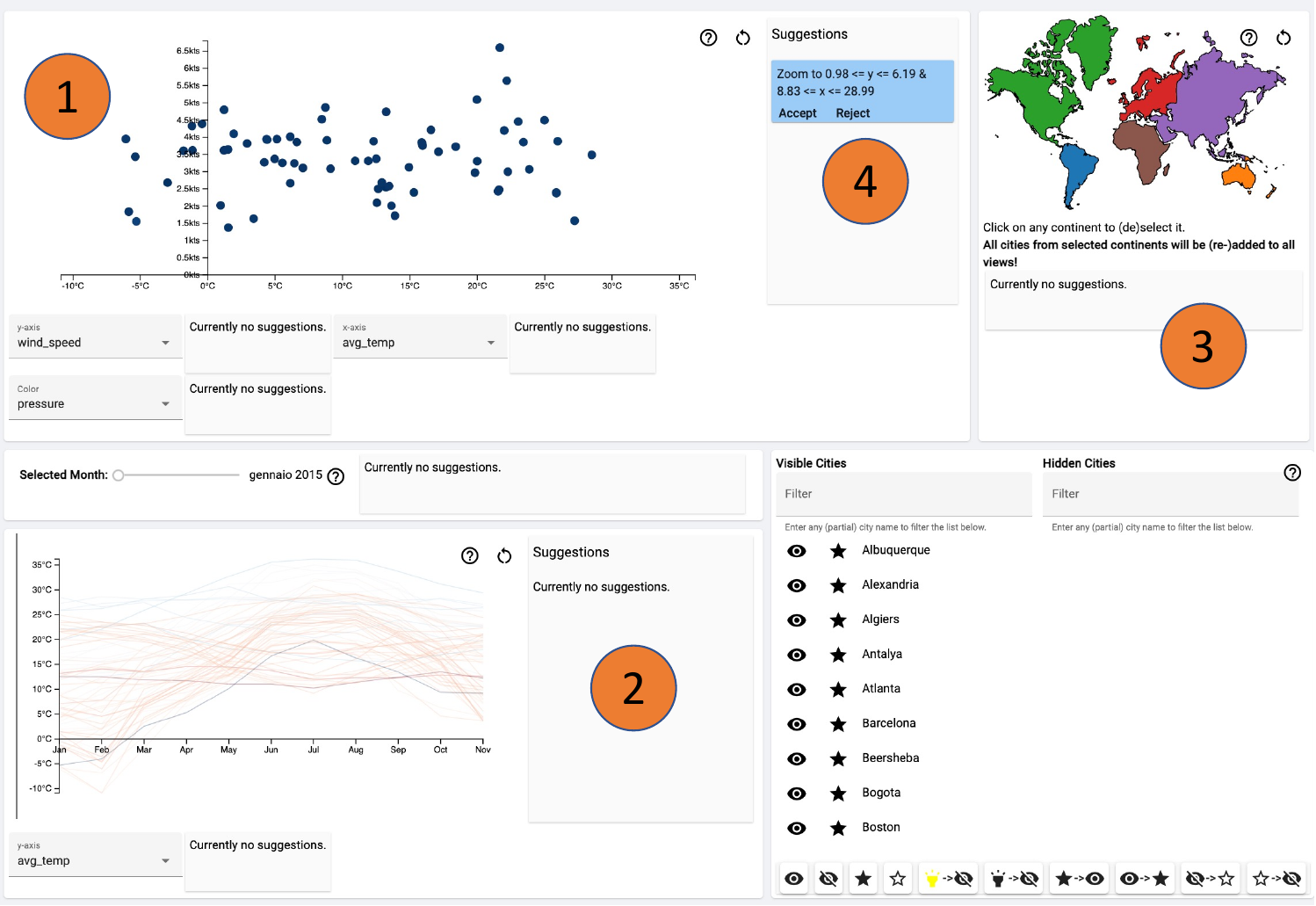}
    \caption{The guidance-enhanced system we used for the expert evaluation. It features different views, including a scatterplot (1), a line chart (2), and a map (3), as well as the possibility to provide guidance to the user. The guidance appears next to each chart as a blue textual box (see for example (4) close to the scatterplot), suggesting to the user what parameters and what data to select in the given visualization.}
    \label{fig:wizard-tool}
\end{figure}

The system employs multiple coordinated views to support various analytical tasks. In our case, it is used to analyze a dataset comprising weather data collected from different cities worldwide. The task of the user is to identify a city with a specific climate given a set of constraints. The data are shown in a scatterplot, a line chart, and a world map, respectively labeled as (1-3) in Fig.~\ref{fig:wizard-tool}. Using these views, the user can choose to visualize, filter, and select different data dimensions and analyze their evolution over time. 
Guidance is provided as blue textual boxes that appear next to the target chart (e.g., see (4) in Fig.~\ref{fig:wizard-tool}) suggesting the user a specific action (e.g., change of axis, value filtering, etc.) on that chart.
By hovering over the suggestion, a preview of its effect is shown to the user.

When the video for the experiment was recorded, guidance could be given directly by a hidden ``wizard'' user, which allowed us to stage different guidance quality criteria from our list using the system features and record it from the user's perspective. The rationale is to make sure that \VAE{} received guidance with specific characteristics so that we could check if and to what extent they could identify them.

All scenarios show the user interacting with the system while solving a task. For the purpose of our study, the guidance was fine-tuned as follows:
\thispagestyle{empty}
{ 
\renewcommand{\theenumi}{F\arabic{enumi}}
\begin{enumerate}[itemsep=-0.1em, topsep=0em, left=0.1em]

\item In all scenarios, guidance is not explained, i.e., no reason is given as to why a suggestion popped up (\textit{un-}explainable guidance).
\item In one scenario, some suggestions were provided with some consistent delay to the user on purpose (\textit{un-}timely guidance).

\item In a different scenario, the system provided irrelevant guidance, i.e., suggestions that were not aligned with the user task (\textit{irrelevant} guidance).
\item In another, the system provided guidance to correct and refine the user selection (\textit{adaptive} guidance).
\item In general, the system could provide, for each view, different degrees of guidance, such as \textit{directing guidance} and \textit{orienting guidance}, as well as no guidance (see the guidance degrees described in~\cite{ceneda2017characterizing}) and the guidance type varied during the analysis (\textit{flexible} guidance).
\end{enumerate}
}
\vspace{-0.1em}

\paragraph{Participants} We contacted 9 \VAE{} by email for the expert evaluation. They are all Ph.D. students or already hold a Ph.D. degree in visualization, and hence can be considered experts. Eventually, 5 \VAE{} participated in our study.

\paragraph{Procedure} The participants received a link to an online survey by email. After signing a consent form, the participants received a first introduction to the analysis tool. They were given a textual description and some static figures of the system and of the different views. Furthermore, we asked them to watch a one-minute video showcasing the main features of the system, how to use them, and the various forms of supported guidance.
After asking if there were questions about the system, we show the video with the six scenarios to the participants in the next phase of the evaluation. The participants could freely watch the video and the different scenarios as often as they wanted. Participants could also go back to the video when answering the questionnaire if they missed some detail. The video lasted 5 minutes; it had no audio but displayed a few captions describing the user's actions or the start of a new scenario.
At the end of the video, we asked the participants to fill out a questionnaire containing the expert heuristics (\VAE{}) listed in Tab.~\ref{table:heuristics}. The participants rated the heuristics following the protocol described in Sect.~\ref{sec:heuristics-use}.

\begin{figure}[htb]
    \centering
    \includegraphics[width=\linewidth]{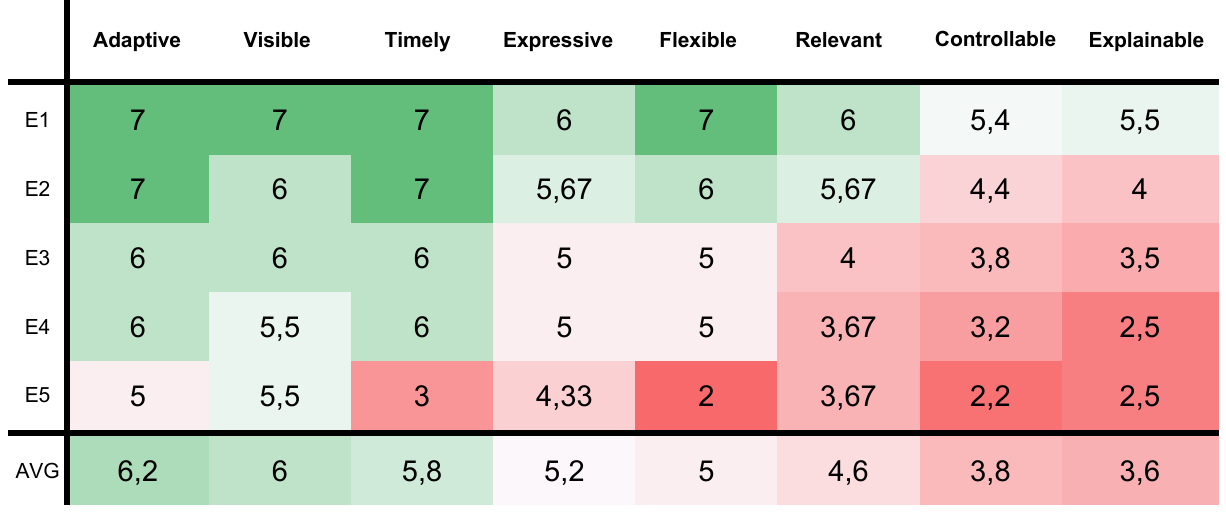}
    \caption{Results of the experts (E1-E5) evaluation. In the table, the ratings are grouped according to the different effectiveness criteria and ordered based on the ratings. Reddish cells represent ratings close to 1 (heuristics violated), while greenish cells represent ratings close to 7 (heuristics applied). Columns are ordered based on the average rating, from higher to lower ratings.}
    \label{fig:table2ndstudy}
\end{figure}

\subsubsection{Outcome of the expert evaluation}
All five experts completed the survey and the results were used for our evaluation.
The results are summarized in Fig.~\ref{fig:table2ndstudy}. For convenience, we grouped the answers according to the quality criteria and calculated an average of the Likert ratings.
The system obtained a total average score of 5.1 (of possible 7). In the context of our evaluation methodology, an average score equal to or above 5 means that the system provides \textit{sufficient} guidance~\cite{wall2018heuristic}.

In general, the expert participants were able to apply our heuristics to identify and rate all the characteristics of the guidance, in line with the features (F1--F5) described earlier. The participants identified the guidance degrees (F5) provided by the system and assessed its \emph{Flexibility}, but also sometimes highlighted the lack of dynamism in adjusting the guidance degree to the user needs. The \emph{Adaptability} of the guidance was positively assessed: The participants recognized that the content of the suggestions was adapted according to the user needs and that the system possibly considered analytical provenance information to generate content. The guidance also received favorable ratings for its \emph{Visibility}, as it was always easy to spot newly generated suggestions. At the same time, \VAE{} found that the system did not provide access (i.e., Visibility) to the guidance parameters and settings. The \emph{Controllability} of the guidance received bad ratings overall. The participants complained that the user was not allowed to ask for guidance, modify its parameters, or turn it off.
Similarly, \VAE{} recognized that the guidance did not have explanations (F1), nor could it be requested by the user, so its rationale or motivation could only be guessed. The \emph{Expressiveness} received an average rating of 5.2, which is greater than five and a weakly satisfactory score. However, the participants highlighted how the encoding of the guidance, which sometimes employed screen coordinates in pixels for the suggestions, was not particularly user-friendly and inexpressive and suggested changing it. The \emph{Timeliness} of the suggestions received good ratings, with one expert recognizing that, in one scenario, the guidance was delayed (F2). Finally, the \emph{Relevance} of the guidance received mixed ratings. Some participants recognized that the guidance was typically on-spot (F4). However, they also recognized the presence of misguidance (F3) and that the missing explanations exacerbated this situation, which made them decrease the final rating.


\subsection{\raisebox{-0.8\mydepth}{\includegraphics[height=1em]{pictures/wraps/blue_icon.png}} \VAU{} Evaluation}

We set up a second study to assess the applicability of our heuristics for \VAU{}. While the goal of the expert evaluation was to assess if a small number of experts could identify a significant number of design issues, our goal for the user evaluation was to assess the coherence of the provided answers and if they could be used to shed light on the effectiveness of a guidance-enhanced system.

\subsubsection{Study design}
As specified in Sect.~\ref{sec:heuristics-use}, we assume users are exposed to our heuristics only after experiencing the guidance to judge its effectiveness. 
For the study, we employed the open-source guidance system 
``Voyager'' (see Fig.~\ref{fig:voyager}), a mixed-initiative system that enables users to perform a guided exploration of a dataset \cite{kanit2016voyager}. 
In terms of guidance, the system offers different kinds of recommendations to support the selection of appropriate visualizations to perform the analysis. Upon the user selecting the data dimensions to analyze, the system generates a list of appropriate visual encodings based on their perceptual properties and statistical measures of the dataset. Moreover, when the user selects a specific visualization, the system can recommend additional data dimensions to be explored as well as alternative encodings of the same data.

\begin{figure}[ht]
    \centering
    \includegraphics[width=\linewidth]{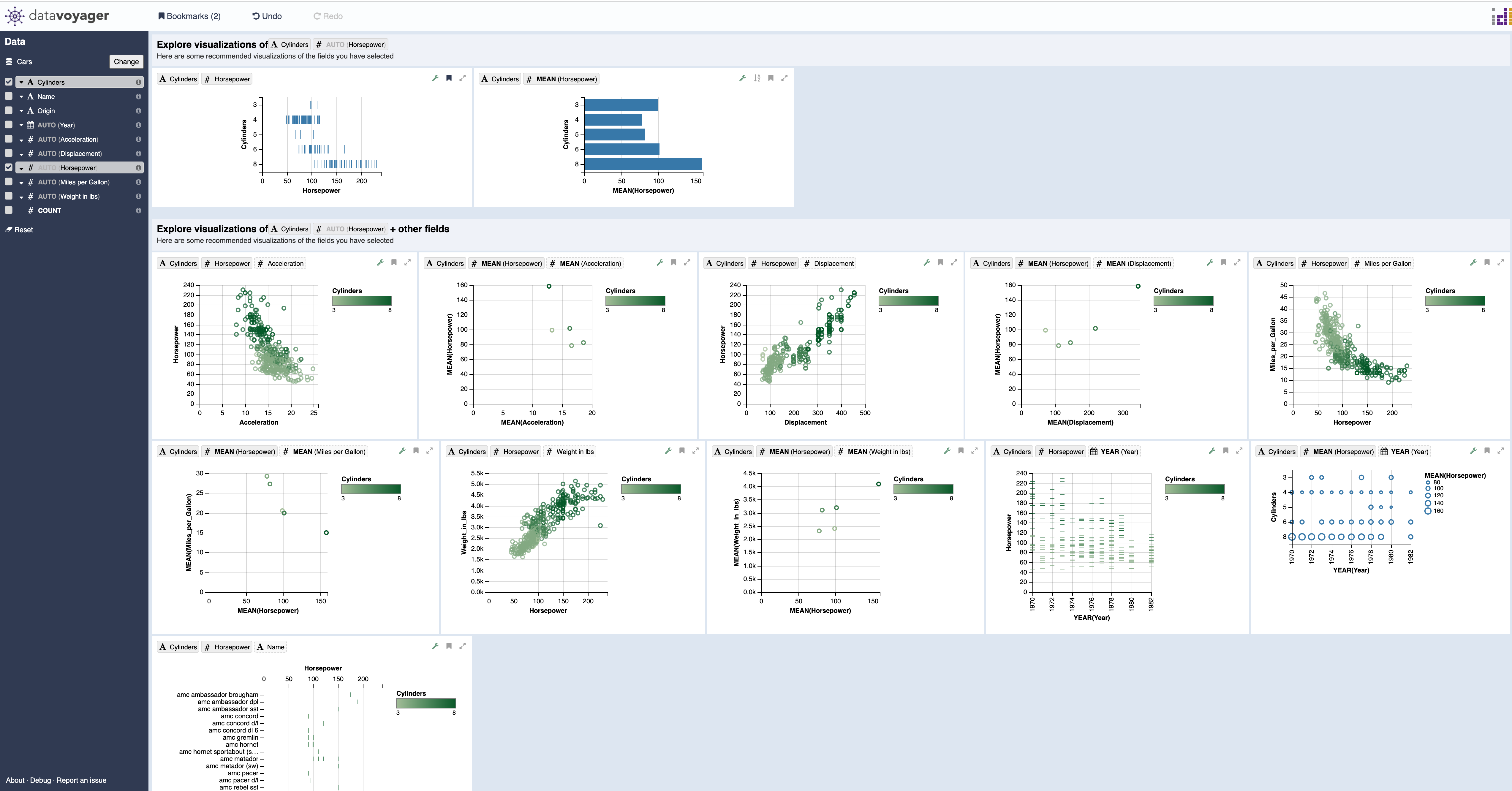}
    \caption{Voyager\cite{kanit2016voyager}, the guidance-enhanced system for our user evaluation.}
    \label{fig:voyager}
\end{figure}

The Voyager system can be categorized as providing \emph{orienting} and \emph{directing} guidance \cite{ceneda2017characterizing}. Orientation is upheld by presenting to the user a broad set of charts to choose from to start the exploration as soon as the data is selected. Directions are given when the user is looking for alternative encodings, to which the system responds by providing a ranked list of different possibilities. Moreover, the system features (to a certain extent) \emph{adaptiveness}, in that the process of generating suggestions considers the user input. Voyager guidance has no \emph{explainability} in its design and implementation, and the timeliness of the recommendations as well as means to control the guidance parameters were not considered during the system design. However, the authors of Voyager found out that the guidance promoted increased coverage of the data variables under analysis (thus, it was deemed \emph{relevant}).

\paragraph{Participants} We asked 50 students to take part in this evaluation. They were part of a bachelor-level course on information design and visualization and have some preliminary knowledge of visualization concepts, but cannot be considered \VAE{}. Students could freely decide to participate either in our study or take part in an alternative assignment. 

\paragraph{Procedure} After signing a consent form, the participants were guided through a practical tutorial to make them familiar with the system features and the guidance, using the embedded \emph{car} dataset. Afterward, they were asked to switch to the embedded \emph{FAA Wildlife Strike} dataset -- which includes statistics of animal striking aircraft -- to solve an analysis task. The task comprised getting familiar with the dataset and ``getting a comprehensive sense of the dataset's content, collecting interesting patterns, trends, or other insights worth sharing with colleagues.'' We did not impose a strict time to complete the task, but we suggested using the tool for 5 minutes at least. To motivate them to interact with the system, we asked them to write a short report (a few lines) of their findings at the end of the study. Finally, after the task was completed, we administered a questionnaire containing our 27 \VAU{} heuristics.


\begin{figure}[hbt]
    \centering
    \includegraphics[width=\linewidth]{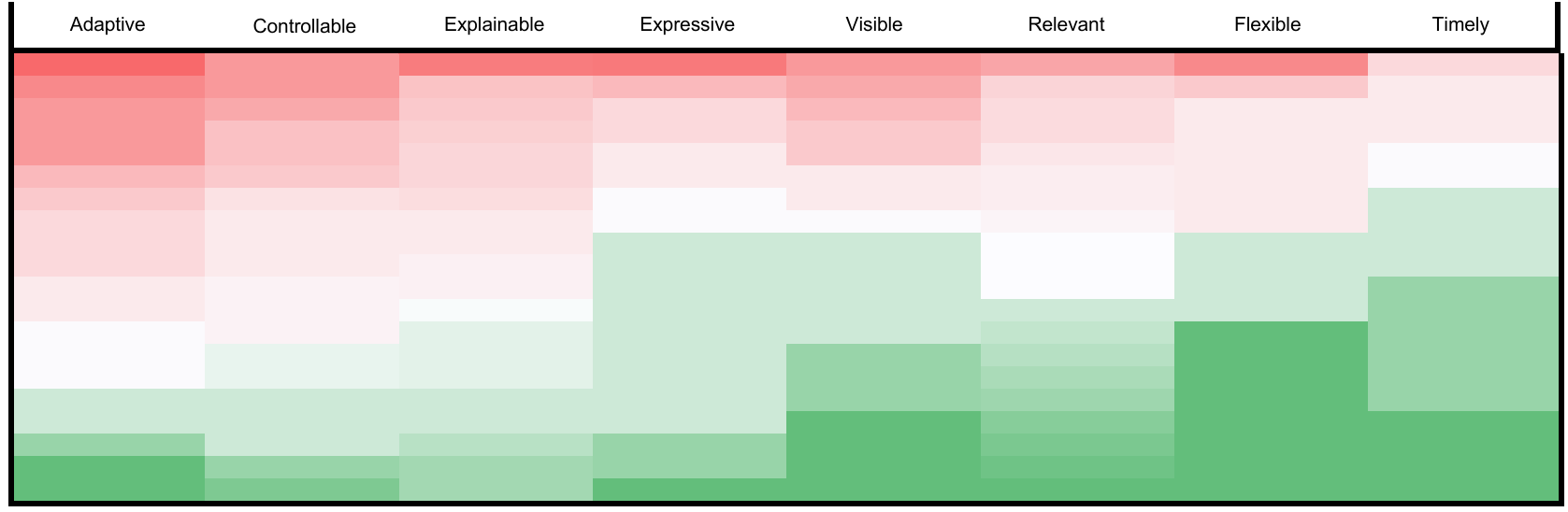}

    \caption{Results of the user evaluation of Voyager~\cite{kanit2016voyager}. For space reasons, we do not report the ratings but simply color-coded the table according to the ratings. Reddish cells represent ratings close to 1 (disagree with heuristics), while greenish cells represent ratings close to 7 (agree with heuristics). Columns are ordered based on the average rating, from lower to higher.}
    \label{fig:table3rdstudy}
\end{figure}

\subsubsection{Outcome of the user evaluation} In total, 39 students answered to our questionnaire. All the received answers were deemed valid (i.e., no incomplete questionnaire was submitted) and valuable for our evaluation.
The results are summarized in Fig.~\ref{fig:table3rdstudy} (the full graph is provided as supplementary material), in which results are grouped according to the quality criteria. To unfold the evaluation outcome, we provide a high-level assessment of these results. Moreover, we conduct a comparative analysis by running the same protocol but with \VAE{} participants, albeit using expert heuristics. 

\paragraph{Qualitative analysis} The results collected provide an assessment of the Voyager system from a guidance perspective. Participants assigned low ratings to the \emph{Adaptability} of the guidance. While the system can provide guidance based on the currently selected data, the participants assessed that the guidance was not adapting to changing needs (UH2, central column, Tab.\ref{table:heuristics}) nor considered past interactions and analytic provenance to update suggestions (UH3). Also, the \emph{Controllability} criterion received mixed evaluations. The participants highlighted that the system does not allow the user to provide feedback to the guidance (UH8) or fine-tune it (UH9). The study participants positively assessed its \emph{Explainability}, as they could easily understand why particular suggestions were provided (UH12). For this reason, they also trusted the guidance (UH13) but highlighted how additional explanations could not be requested on demand (UH14).
Finally, analyzing the \emph{Relevance} of the guidance, the participants reported that they were not supported in getting out of problematic situations nor was the possibility of making errors reduced significantly. This outcome is in line with the tool specifications~\cite{kanit2016voyager}. After all, the user is always in charge of making decisions. The system cannot determine if errors are done when solving a task, as it only suggests how to visualize the data appropriately. However, the students reported that the guidance was appropriate for their task and that it helped them reason about the data (UH23--24).

\paragraph{Quantitative Analysis} \textcolor{black}{Following standard practices for developing and assessing surveys and heuristics\cite{hinkin1997scale}, we performed a quantitative analysis of the results obtained, according to the different quality criteria. In particular, we applied Cronbach's alpha\cite{cronbach1951coefficient} to the results obtained. This is a measure of reliability used to calculate the internal consistency of tests and measures (as our heuristics). We obtained an average score of $alpha=0.87$, indicating that the collected responses are consistent. This consistency suggests that our heuristics are reliable and that our criteria are not redundant.}


\subsubsection{Comparative analysis}
\paragraph{Comparison with original evaluation} We provide a comparative analysis of the user evaluation results, examining them in contrast to the results of the original evaluation performed by the authors of Voyager\cite{kanit2016voyager}.
Our evaluation approach provides a much more detailed assessment of the guidance compared to how the system was originally evaluated. The initial evaluation \cite{kanit2016voyager} aimed to discover the value of adding guidance and its usefulness in supporting data exploration (i.e., get an overview). The results of the initial evaluation partially cover some aspects, which in our methodology are considered by the criterion of \emph{Relevance}. The authors also assessed the users' trust towards the recommendations (the same aspect is part of our \emph{Explainability} criteria, which has a broader scope, though). In our evaluation, \VAU{} reported, as in the original evaluation, that they trusted the system recommendations. However, our methodology expands such assessment showing a need for more adaptation, flexibility, suitable explanations, and control over the guidance (as shown in  Fig.\ref{fig:table3rdstudy}).

\paragraph{Comparison with \VAE{} assessment}  We also ran a parallel expert evaluation of Voyager, using the \VAE{} heuristics and involving three additional visualization and guidance experts. The results are shown in Figure~\ref{fig:table4thstudy}. In summary, by comparing results with Figure~\ref{fig:table3rdstudy}, we can see how the results are consistent between the two groups of evaluators. However, we can also see how the results of some criteria slightly diverge between \VAU{} and \VAE{} on some aspects which we describe in the following.
Both groups of evaluators reported, for instance, that the guidance was \emph{Timely}, and both assigned low ratings to its \emph{Controllability} and \emph{Explainability}. However, \VAU{} assigned low ratings to the guidance \emph{Adapatability}, while \VAE{}, on average, provided higher ratings, signaling that the design of the guidance could be potentially improved. Still, on the \emph{Adapatability} criterion, only one \VAE{} (1 of 3) recognized that the content of the suggestions does not consider analytic provenance or user preferences, rating it 2. On the other hand, \VAU{} rated positively the \emph{Relevance} of the guidance in supporting dataset exploration and getting an overview. At the same time, \VAE{} consistently highlighted that the system could not recognize and prevent errors or biases.

\begin{figure}[htb]
    \centering
    \includegraphics[width=\linewidth]{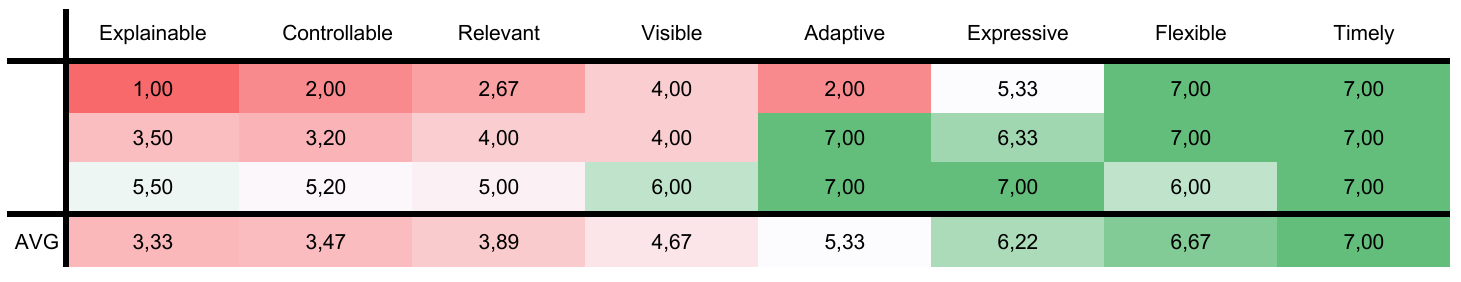}

    \caption{Results of the expert evaluation of Voyager~\cite{kanit2016voyager}. Reddish cells represent ratings close to 1 (heuristics violated), while greenish cells represent ratings close to 7 (heuristics applied). Columns are ordered based on the average rating, from lower to higher.}
    \label{fig:table4thstudy}
\end{figure}

\section{Discussion and Lessons Learned} 

\paragraph{Importance of quality criteria} The results show that the effectiveness of guidance comes from a combination of different characteristics, which we match to our eight quality criteria. The majority of the experts we consulted characterized the effectiveness of guidance in terms of its ability to support the completion of tasks (using terms such as \emph{meaningful}, \emph{facilitating}, \emph{goal-oriented}), its transparency to the user (using terms as \emph{clear}, \emph{transparent}, \emph{descriptive}) and its adaptability to the analysis context (with terms such as \emph{user-/task-dependant}, \emph{context-sensitive}). These characteristics can be expressed in terms of the \emph{Relevance}, the \emph{Explainability}, and the \emph{Adaptiveness} of guidance content.
Whereas we rated and assigned the same weight to all the criteria, these results suggest that not all of them have the same impact on the final guidance design, and depending on the context some could be prioritized over others.
These results provide valuable insights 
on how users perceive the guidance qualities and therefore support this kind of strategic decision.

\paragraph{\VAE{}  must be really experts} During the design of the evaluation framework, we ran internal tests before running the real evaluations. We encountered some problems during the expert trials, mainly accounting for the lack of knowledge of the guidance concepts. In particular, as mentioned, we phrased the \VAE{} heuristics using guidance terminology used commonly in the literature. During the tests, we noticed that the two \VAE{} involved were not entirely familiar with the terminology, which led them to misinterpret some of the heuristics. Although it seems obvious and expected that expert heuristics should be assessed only by real experts, we want to highlight how easy it is to make mistakes and obtain unreliable results. A simple solution to this problem could be rephrasing the heuristics or providing the necessary information about the terminology used before running the evaluation. This issue also raises a point about the necessity to strengthen the dissemination of guidance literature in our community.

\paragraph{Completeness} \textcolor{black}{We combined a bottom-up (i.e., reviewing the literature) and a top-down (i.e., input from the VA scholars) approach for the generation of the criteria, aiming for complete coverage of desirable guidance properties. However, we acknowledge that more work is potentially needed to assess the exhaustiveness of our list.}
Valuable input for this process could derive from the application of our heuristics in practical scenarios, beyond the initial assessment described in this paper.}

\paragraph{Templates for experts and user testing} Finally, we want to highlight that the main contribution of this paper lies not only in the description and discussion of a methodology to evaluate guidance approaches in VA. In addition, we provide a protocol (Sect.~\ref{sec:heuristics-use}) for their application in practical scenarios, for rating and interpreting results, enrolling participants, as well as templates of questionnaires to be used by experts and users for testing in this Git repository\cite{gitgit}.

\section{Conclusion}
This paper identifies the crucial factors contributing to the value and effectiveness of guidance. Similar to the value of visualizations \cite{stasko2014value}, designing and developing effective guidance depends on balancing different system features.
We identified eight quality criteria, which we instantiated into two sets of heuristics to evaluate and define effective guidance. We ran two evaluations to assess how \VAE{} and \VAU{} used our methodology in practice. For the expert evaluation, we created an ad-hoc system with specific guidance features. The results show that the experts were able to coherently identify the strengths and weaknesses of the proposed guidance, showing how our methodology can effectively contribute to guidance design already from its early development stages. We employed a well-known guidance-enhanced visualization tool (Voyager) for user evaluation, asked some \VAU{} to use it, and evaluated the provided support. \VAU{} could rate the guidance features consistently, showing a convergence of their and experts' ratings. We show the value of our methodology by providing an in-depth assessment of the provided suggestions.


\acknowledgments{
This work was funded by Vienna Science and Technology Fund (WWTF) under grant [10.47379/ICT19047] and by DoRIAH (FFG Grant \#880883).
We also want to thank Markus B\"ogl and Emily Wall for their support with the evaluation, and the anonymous reviewers for their constructive feedback.
}

\bibliographystyle{abbrv-doi-hyperref}

\bibliography{template}

\begin{thebibliography}{10}

\bibitem{gitgit}
{Templates and Protocol}.
\newblock
  \url{https://gitlab.cvast.tuwien.ac.at/dceneda/heuristics-guidance-effectiveness/}.
\newblock [Online; accessed 31-March-2023].

\bibitem{amar2004best}
R.~Amar and J.~Stasko.
\newblock A knowledge task-based framework for design and evaluation of
  information visualizations.
\newblock In {\em Proc. IEEE Symp. on Information Visualization}, pp. 143--150.
  IEEE, 2004. \href{https://doi.org/10.1109/INFVIS.2004.10}
{doi: {{%
10\hspace{.1pt}\discretionary{.}{%
}{.}\hspace{.4pt}1109\discretionary{/}{%
}{/}INFVIS\hspace{.1pt}\discretionary{.}{%
}{.}\hspace{.4pt}2004\hspace{.1pt}\discretionary{.}{%
}{.}\hspace{.4pt}10}}}


\bibitem{BERGERON1972175}
R.~Bergeron, J.~Gannon, D.~Shecter, F.~Tompa, and A.~V. Dam.
\newblock Systems programming languages.
\newblock In M.~Rubtnoff, ed., {\em Advances in Computers}, vol.~12, pp.
  175--284. Elsevier, 1972.
  \href{https://doi.org/10.1016/S0065-2458(08)60510-0}
{doi: {{%
10\hspace{.1pt}\discretionary{.}{%
}{.}\hspace{.4pt}1016\discretionary{/}{%
}{/}S0065\discretionary{%
}{-}{-}2458\discretionary{%
}{(}{(}08\discretionary{)}{%
}{)}60510\discretionary{%
}{-}{-}0}}}


\bibitem{bouali2016vizassist}
F.~Bouali, A.~Guettala, and G.~Venturini.
\newblock {VizAssist}: An interactive user assistant for visual data mining.
\newblock {\em The Visual Computer}, 32(11):1447--1463, 2016.
  \href{https://doi.org/10.1007/s00371-015-1132-9}
{doi: {{%
10\hspace{.1pt}\discretionary{.}{%
}{.}\hspace{.4pt}1007\discretionary{/}{%
}{/}s00371\discretionary{%
}{-}{-}015\discretionary{%
}{-}{-}1132\discretionary{%
}{-}{-}9}}}


\bibitem{ceneda2020guide}
D.~Ceneda, N.~Andrienko, G.~Andrienko, T.~Gschwandtner, S.~Miksch,
  N.~Piccolotto, T.~Schreck, M.~Streit, J.~Suschnigg, and C.~Tominski.
\newblock Guide me in analysis: A framework for guidance designers.
\newblock {\em Computer Graphics Forum}, 39(6):269--288, 2020.
  \href{https://doi.org/10.1111/cgf.14017}
{doi: {{%
10\hspace{.1pt}\discretionary{.}{%
}{.}\hspace{.4pt}1111\discretionary{/}{%
}{/}cgf\hspace{.1pt}\discretionary{.}{%
}{.}\hspace{.4pt}14017}}}


\bibitem{ceneda2021show}
D.~Ceneda, A.~Arleo, T.~Gschwandtner, and S.~Miksch.
\newblock Show me your face: Towards an automated method to provide timely
  guidance in visual analytics.
\newblock {\em IEEE Trans. Visualization and Computer Graphics},
  28(12):4570--4581, 2021. \href{https://doi.org/10.1109/TVCG.2021.3094870}
{doi: {{%
10\hspace{.1pt}\discretionary{.}{%
}{.}\hspace{.4pt}1109\discretionary{/}{%
}{/}TVCG\hspace{.1pt}\discretionary{.}{%
}{.}\hspace{.4pt}2021\hspace{.1pt}\discretionary{.}{%
}{.}\hspace{.4pt}3094870}}}


\bibitem{ceneda2017characterizing}
D.~Ceneda, T.~Gschwandtner, T.~May, S.~Miksch, H.-J. Schulz, M.~Streit, and
  C.~Tominski.
\newblock Characterizing guidance in visual analytics.
\newblock {\em IEEE Trans. Visualization and Computer Graphics},
  23(1):111--120, 2017. \href{https://doi.org/10.1109/TVCG.2016.2598468}
{doi: {{%
10\hspace{.1pt}\discretionary{.}{%
}{.}\hspace{.4pt}1109\discretionary{/}{%
}{/}TVCG\hspace{.1pt}\discretionary{.}{%
}{.}\hspace{.4pt}2016\hspace{.1pt}\discretionary{.}{%
}{.}\hspace{.4pt}2598468}}}


\bibitem{ceneda2018decision}
D.~Ceneda, T.~Gschwandtner, T.~May, S.~Miksch, M.~Streit, and C.~Tominski.
\newblock Guidance or no guidance? {A} decision tree can help.
\newblock In {\em EuroVA: International Workshop on Visual Analytics}, pp.
  19--23. Eurographics Digital Library, 2018.
  \href{https://doi.org/10.2312/eurova.20181107}
{doi: {{%
10\hspace{.1pt}\discretionary{.}{%
}{.}\hspace{.4pt}2312\discretionary{/}{%
}{/}eurova\hspace{.1pt}\discretionary{.}{%
}{.}\hspace{.4pt}20181107}}}


\bibitem{ceneda2019review}
D.~Ceneda, T.~Gschwandtner, and S.~Miksch.
\newblock A review of guidance approaches in visual data analysis: A multifocal
  perspective.
\newblock {\em Computer Graphics Forum}, 38(3):861--879, 2019.
  \href{https://doi.org/10.1111/cgf.13730}
{doi: {{%
10\hspace{.1pt}\discretionary{.}{%
}{.}\hspace{.4pt}1111\discretionary{/}{%
}{/}cgf\hspace{.1pt}\discretionary{.}{%
}{.}\hspace{.4pt}13730}}}


\bibitem{ceneda2018guided}
D.~Ceneda, T.~Gschwandtner, S.~Miksch, and C.~Tominski.
\newblock Guided visual exploration of cyclical patterns in time-series.
\newblock In T.~M{\"o}ller and S.~Liu, eds., {\em Visualization in Data Science
  (VDS)}. IEEE Computer Society, 2018.

\bibitem{chen2000empirical}
C.~Chen and M.~P. Czerwinski.
\newblock Empirical evaluation of information visualizations: An introduction.
\newblock {\em Int. J. of Human-Computer Studies}, 53(5):631--635, 2000.
  \href{https://doi.org/10.1006/ijhc.2000.0421}
{doi: {{%
10\hspace{.1pt}\discretionary{.}{%
}{.}\hspace{.4pt}1006\discretionary{/}{%
}{/}ijhc\hspace{.1pt}\discretionary{.}{%
}{.}\hspace{.4pt}2000\hspace{.1pt}\discretionary{.}{%
}{.}\hspace{.4pt}0421}}}


\bibitem{chen2000empirical1}
C.~Chen and Y.~Yu.
\newblock Empirical studies of information visualization: A meta-analysis.
\newblock {\em Int. J. of Human-Computer Studies}, 53(5):851--866, 2000.
  \href{https://doi.org/10.1006/ijhc.2000.0422}
{doi: {{%
10\hspace{.1pt}\discretionary{.}{%
}{.}\hspace{.4pt}1006\discretionary{/}{%
}{/}ijhc\hspace{.1pt}\discretionary{.}{%
}{.}\hspace{.4pt}2000\hspace{.1pt}\discretionary{.}{%
}{.}\hspace{.4pt}0422}}}


\bibitem{choo2010ivisclassifier}
J.~Choo, H.~Lee, J.~Kihm, and H.~Park.
\newblock {iVisClassifier}: An interactive visual analytics system for
  classification based on supervised dimension reduction.
\newblock In {\em Proc. IEEE Symp. on Visual Analytics Science and Technology},
  pp. 27--34. IEEE, 2010. \href{https://doi.org/10.1109/VAST.2010.5652443}
{doi: {{%
10\hspace{.1pt}\discretionary{.}{%
}{.}\hspace{.4pt}1109\discretionary{/}{%
}{/}VAST\hspace{.1pt}\discretionary{.}{%
}{.}\hspace{.4pt}2010\hspace{.1pt}\discretionary{.}{%
}{.}\hspace{.4pt}5652443}}}


\bibitem{collins2018guidance}
C.~Collins, N.~Andrienko, T.~Schreck, J.~Yang, J.~Choo, U.~Engelke, A.~Jena,
  and T.~Dwyer.
\newblock Guidance in the human--machine analytics process.
\newblock {\em Visual Informatics}, 2(3):166--180, 2018.
  \href{https://doi.org/10.1016/j.visinf.2018.09.003}
{doi: {{%
10\hspace{.1pt}\discretionary{.}{%
}{.}\hspace{.4pt}1016\discretionary{/}{%
}{/}j\hspace{.1pt}\discretionary{.}{%
}{.}\hspace{.4pt}visinf\hspace{.1pt}\discretionary{.}{%
}{.}\hspace{.4pt}2018\hspace{.1pt}\discretionary{.}{%
}{.}\hspace{.4pt}09\hspace{.1pt}\discretionary{.}{%
}{.}\hspace{.4pt}003}}}


\bibitem{cook2005illuminating}
K.~A. Cook and J.~J. Thomas.
\newblock Illuminating the path: The research and development agenda for visual
  analytics.
\newblock Technical report, Pacific Northwest National Lab (PNNL), Richland, WA
  (United States), 2005.

\bibitem{cronbach1951coefficient}
L.~J. Cronbach.
\newblock Coefficient alpha and the internal structure of tests.
\newblock {\em psychometrika}, 16(3):297--334, 1951.
  \href{https://doi.org/10.1007/BF02310555}
{doi: {{%
10\hspace{.1pt}\discretionary{.}{%
}{.}\hspace{.4pt}1007\discretionary{/}{%
}{/}BF02310555}}}


\bibitem{forsell2012evaluation}
C.~Forsell.
\newblock Evaluation in information visualization: Heuristic evaluation.
\newblock In {\em Proc. IEEE Symp. on Information Visualization}, pp. 136--142.
  IEEE, 2012. \href{https://doi.org/10.1109/IV.2012.33}
{doi: {{%
10\hspace{.1pt}\discretionary{.}{%
}{.}\hspace{.4pt}1109\discretionary{/}{%
}{/}IV\hspace{.1pt}\discretionary{.}{%
}{.}\hspace{.4pt}2012\hspace{.1pt}\discretionary{.}{%
}{.}\hspace{.4pt}33}}}


\bibitem{forsell2010heuristic}
C.~Forsell and J.~Johansson.
\newblock An heuristic set for evaluation in information visualization.
\newblock In {\em Proc. Int. Conf. on Advanced Visual Interfaces}, pp.
  199--206, 2010. \href{https://doi.org/10.1145/1842993.1843029}
{doi: {{%
10\hspace{.1pt}\discretionary{.}{%
}{.}\hspace{.4pt}1145\discretionary{/}{%
}{/}1842993\hspace{.1pt}\discretionary{.}{%
}{.}\hspace{.4pt}1843029}}}


\bibitem{gladisch2013navigation}
S.~Gladisch, H.~Schumann, and C.~Tominski.
\newblock Navigation recommendations for exploring hierarchical graphs.
\newblock In {\em Int. Symp. on Visual Computing}, pp. 36--47. Springer, 2013.
  \href{https://doi.org/10.1007/978-3-642-41939-3_4}
{doi: {{%
10\hspace{.1pt}\discretionary{.}{%
}{.}\hspace{.4pt}1007\discretionary{/}{%
}{/}978\discretionary{%
}{-}{-}3\discretionary{%
}{-}{-}642\discretionary{%
}{-}{-}41939\discretionary{%
}{-}{-}3\_4}}}


\bibitem{gotz2010harvest}
D.~Gotz, J.~Lu, P.~Kissa, N.~Cao, W.~H. Qian, S.~X. Liu, and M.~X. Zhou.
\newblock {HARVEST}: An intelligent visual analytic tool for the masses.
\newblock In {\em Proc. of Int. Workshop on Intelligent Visual Interfaces for
  Text Analysis}, pp. 1--4. ACM, 2010.
  \href{https://doi.org/10.1145/2002353.2002355}
{doi: {{%
10\hspace{.1pt}\discretionary{.}{%
}{.}\hspace{.4pt}1145\discretionary{/}{%
}{/}2002353\hspace{.1pt}\discretionary{.}{%
}{.}\hspace{.4pt}2002355}}}


\bibitem{gotz2009behavior}
D.~Gotz and Z.~Wen.
\newblock Behavior-driven visualization recommendation.
\newblock In {\em Proc. of ACM Conf. on Intelligent User Interfaces}, pp.
  315--324. ACM, 2009. \href{https://doi.org/10.1145/1502650.1502695}
{doi: {{%
10\hspace{.1pt}\discretionary{.}{%
}{.}\hspace{.4pt}1145\discretionary{/}{%
}{/}1502650\hspace{.1pt}\discretionary{.}{%
}{.}\hspace{.4pt}1502695}}}


\bibitem{Han23Guidance}
W.~Han and H.-J. Schulz.
\newblock Providing visual analytics guidance through decision support.
\newblock {\em Information Visualization}, 22(2):140--165, 2023.
  \href{https://doi.org/10.1177/14738716221147289}
{doi: {{%
10\hspace{.1pt}\discretionary{.}{%
}{.}\hspace{.4pt}1177\discretionary{/}{%
}{/}14738716221147289}}}


\bibitem{he2022beauvis}
T.~He, P.~Isenberg, R.~Dachselt, and T.~Isenberg.
\newblock {BeauVis}: A validated scale for measuring the aesthetic pleasure of
  visual representations.
\newblock {\em IEEE Trans. Visualization and Computer Graphics},
  29(1):363--373, 2022. \href{https://doi.org/10.1109/TVCG.2022.3209390}
{doi: {{%
10\hspace{.1pt}\discretionary{.}{%
}{.}\hspace{.4pt}1109\discretionary{/}{%
}{/}TVCG\hspace{.1pt}\discretionary{.}{%
}{.}\hspace{.4pt}2022\hspace{.1pt}\discretionary{.}{%
}{.}\hspace{.4pt}3209390}}}


\bibitem{heer2005vizster}
J.~Heer and D.~Boyd.
\newblock Vizster: Visualizing online social networks.
\newblock In {\em Proc. IEEE Symp. on Information Visualization}, pp. 32--39.
  IEEE, 2005. \href{https://doi.org/10.1109/INFVIS.2005.1532126}
{doi: {{%
10\hspace{.1pt}\discretionary{.}{%
}{.}\hspace{.4pt}1109\discretionary{/}{%
}{/}INFVIS\hspace{.1pt}\discretionary{.}{%
}{.}\hspace{.4pt}2005\hspace{.1pt}\discretionary{.}{%
}{.}\hspace{.4pt}1532126}}}


\bibitem{heer2007animated}
J.~Heer and G.~Robertson.
\newblock Animated transitions in statistical data graphics.
\newblock {\em IEEE Trans. Visualization and Computer Graphics},
  13(6):1240--1247, 2007. \href{https://doi.org/10.1109/TVCG.2007.70539}
{doi: {{%
10\hspace{.1pt}\discretionary{.}{%
}{.}\hspace{.4pt}1109\discretionary{/}{%
}{/}TVCG\hspace{.1pt}\discretionary{.}{%
}{.}\hspace{.4pt}2007\hspace{.1pt}\discretionary{.}{%
}{.}\hspace{.4pt}70539}}}


\bibitem{hinkin1997scale}
T.~R. Hinkin, J.~B. Tracey, and C.~A. Enz.
\newblock Scale construction: Developing reliable and valid measurement
  instruments.
\newblock {\em Journal of Hospitality \& Tourism Research}, 21(1):100--120,
  1997. \href{https://doi.org/10.1177/109634809702100108}
{doi: {{%
10\hspace{.1pt}\discretionary{.}{%
}{.}\hspace{.4pt}1177\discretionary{/}{%
}{/}109634809702100108}}}


\bibitem{kandel2011wrangler}
S.~Kandel, A.~Paepcke, J.~Hellerstein, and J.~Heer.
\newblock Wrangler: Interactive visual specification of data transformation
  scripts.
\newblock In {\em Proc. SIGCHI Conf. on Human Factors in Computing Systems},
  pp. 3363--3372. ACM, 2011. \href{https://doi.org/10.1145/1978942.1979444}
{doi: {{%
10\hspace{.1pt}\discretionary{.}{%
}{.}\hspace{.4pt}1145\discretionary{/}{%
}{/}1978942\hspace{.1pt}\discretionary{.}{%
}{.}\hspace{.4pt}1979444}}}


\bibitem{kandel2012profiler}
S.~Kandel, R.~Parikh, A.~Paepcke, J.~M. Hellerstein, and J.~Heer.
\newblock Profiler: Integrated statistical analysis and visualization for data
  quality assessment.
\newblock In {\em Proc. Int. Conf. on Advanced Visual Interfaces}, pp.
  547--554. ACM, 2012. \href{https://doi.org/10.1145/2254556.2254659}
{doi: {{%
10\hspace{.1pt}\discretionary{.}{%
}{.}\hspace{.4pt}1145\discretionary{/}{%
}{/}2254556\hspace{.1pt}\discretionary{.}{%
}{.}\hspace{.4pt}2254659}}}


\bibitem{keim2008visual}
D.~A. Keim, F.~Mansmann, J.~Schneidewind, J.~Thomas, and H.~Ziegler.
\newblock Visual analytics: Scope and challenges.
\newblock In {\em Visual Data Mining}, pp. 76--90. Springer, 2008.
  \href{https://doi.org/10.1007/978-3-540-71080-6_6}
{doi: {{%
10\hspace{.1pt}\discretionary{.}{%
}{.}\hspace{.4pt}1007\discretionary{/}{%
}{/}978\discretionary{%
}{-}{-}3\discretionary{%
}{-}{-}540\discretionary{%
}{-}{-}71080\discretionary{%
}{-}{-}6\_6}}}


\bibitem{krause2016interacting}
J.~Krause, A.~Perer, and K.~Ng.
\newblock Interacting with predictions: Visual inspection of black-box machine
  learning models.
\newblock In {\em Proc. SIGCHI Conf. on Human Factors in Computing Systems},
  pp. 5686--5697. ACM, 2016. \href{https://doi.org/10.1145/2858036.2858529}
{doi: {{%
10\hspace{.1pt}\discretionary{.}{%
}{.}\hspace{.4pt}1145\discretionary{/}{%
}{/}2858036\hspace{.1pt}\discretionary{.}{%
}{.}\hspace{.4pt}2858529}}}


\bibitem{krishnamoorthy2006personalized}
G.~Krishnamoorthy and P.~Brusilovsky.
\newblock Personalized guidance for example selection in an explanatory
  visualization system.
\newblock In {\em Proc. of E-Learn: World Conf. on E-Learning in Corporate,
  Government, Healthcare, and Higher Education}, pp. 2122--2127. AACE, 2006.

\bibitem{luboschikheterogeneity}
M.~Luboschik, C.~Maus, H.-J. Schulz, H.~Schumann, and A.~Uhrmacher.
\newblock Heterogeneity-based guidance for exploring multiscale data in systems
  biology.
\newblock In {\em Proc. IEEE Symp. on Biological Data Visualization}, pp.
  33--40. IEEE, 2012. \href{https://doi.org/10.1109/BioVis.2012.6378590}
{doi: {{%
10\hspace{.1pt}\discretionary{.}{%
}{.}\hspace{.4pt}1109\discretionary{/}{%
}{/}BioVis\hspace{.1pt}\discretionary{.}{%
}{.}\hspace{.4pt}2012\hspace{.1pt}\discretionary{.}{%
}{.}\hspace{.4pt}6378590}}}


\bibitem{mackinlay1986automating}
J.~Mackinlay.
\newblock Automating the design of graphical presentations of relational
  information.
\newblock {\em ACM Trans. on Graphics (Tog)}, 5(2):110--141, 1986.
  \href{https://doi.org/10.1145/22949.22950}
{doi: {{%
10\hspace{.1pt}\discretionary{.}{%
}{.}\hspace{.4pt}1145\discretionary{/}{%
}{/}22949\hspace{.1pt}\discretionary{.}{%
}{.}\hspace{.4pt}22950}}}


\bibitem{mankoff2003heuristic}
J.~Mankoff, A.~K. Dey, G.~Hsieh, J.~Kientz, S.~Lederer, and M.~Ames.
\newblock Heuristic evaluation of ambient displays.
\newblock In {\em Proc. SIGCHI Conf. on Human Factors in Computing Systems},
  pp. 169--176, 2003. \href{https://doi.org/10.1145/642611.642642}
{doi: {{%
10\hspace{.1pt}\discretionary{.}{%
}{.}\hspace{.4pt}1145\discretionary{/}{%
}{/}642611\hspace{.1pt}\discretionary{.}{%
}{.}\hspace{.4pt}642642}}}


\bibitem{may2011guiding}
T.~May, A.~Bannach, J.~Davey, T.~Ruppert, and J.~Kohlhammer.
\newblock Guiding feature subset selection with an interactive visualization.
\newblock In {\em Proc. IEEE Symp. on Visual Analytics Science and Technology},
  pp. 111--120. IEEE, 2011. \href{https://doi.org/10.1109/VAST.2011.6102448}
{doi: {{%
10\hspace{.1pt}\discretionary{.}{%
}{.}\hspace{.4pt}1109\discretionary{/}{%
}{/}VAST\hspace{.1pt}\discretionary{.}{%
}{.}\hspace{.4pt}2011\hspace{.1pt}\discretionary{.}{%
}{.}\hspace{.4pt}6102448}}}


\bibitem{miksch2014matter}
S.~Miksch and W.~Aigner.
\newblock A matter of time: Applying a data--users--tasks design triangle to
  visual analytics of time-oriented data.
\newblock {\em Computers \& Graphics}, 38:286--290, 2014.
  \href{https://doi.org/10.1016/j.cag.2013.11.002}
{doi: {{%
10\hspace{.1pt}\discretionary{.}{%
}{.}\hspace{.4pt}1016\discretionary{/}{%
}{/}j\hspace{.1pt}\discretionary{.}{%
}{.}\hspace{.4pt}cag\hspace{.1pt}\discretionary{.}{%
}{.}\hspace{.4pt}2013\hspace{.1pt}\discretionary{.}{%
}{.}\hspace{.4pt}11\hspace{.1pt}\discretionary{.}{%
}{.}\hspace{.4pt}002}}}


\bibitem{molich1990improving}
R.~Molich and J.~Nielsen.
\newblock Improving a human-computer dialogue.
\newblock {\em Communications of the ACM}, 33(3):338--348, 1990.
  \href{https://doi.org/10.1145/77481.77486}
{doi: {{%
10\hspace{.1pt}\discretionary{.}{%
}{.}\hspace{.4pt}1145\discretionary{/}{%
}{/}77481\hspace{.1pt}\discretionary{.}{%
}{.}\hspace{.4pt}77486}}}


\bibitem{nielsen2005ten}
J.~Nielsen.
\newblock Heuristic evaluation.
\newblock {\em Usability Inspection Mehods}, 1994.
  \href{https://doi.org/10.5555/189200.189209}
{doi: {{%
10\hspace{.1pt}\discretionary{.}{%
}{.}\hspace{.4pt}5555\discretionary{/}{%
}{/}189200\hspace{.1pt}\discretionary{.}{%
}{.}\hspace{.4pt}189209}}}


\bibitem{o2015designscape}
P.~O'Donovan, A.~Agarwala, and A.~Hertzmann.
\newblock {DesignScape}: Design with interactive layout suggestions.
\newblock In {\em Proc. SIGCHI Conf. on Human Factors in Computing Systems},
  pp. 1221--1224. ACM, 2015. \href{https://doi.org/10.1145/2702123.2702149}
{doi: {{%
10\hspace{.1pt}\discretionary{.}{%
}{.}\hspace{.4pt}1145\discretionary{/}{%
}{/}2702123\hspace{.1pt}\discretionary{.}{%
}{.}\hspace{.4pt}2702149}}}


\bibitem{perez2022typology}
I.~P{\'e}rez-Messina, D.~Ceneda, M.~El-Assady, S.~Miksch, and F.~Sperrle.
\newblock A typology of guidance tasks in mixed-initiative visual analytics
  environments.
\newblock In {\em Computer Graphics Forum}, vol.~41, pp. 465--476. Wiley, 2022.
  \href{https://doi.org/10.1111/cgf.14555}
{doi: {{%
10\hspace{.1pt}\discretionary{.}{%
}{.}\hspace{.4pt}1111\discretionary{/}{%
}{/}cgf\hspace{.1pt}\discretionary{.}{%
}{.}\hspace{.4pt}14555}}}


\bibitem{plaisant2004challenge}
C.~Plaisant.
\newblock The challenge of information visualization evaluation.
\newblock In {\em Proc. Int. Conf. on Advanced Visual Interfaces}, pp.
  109--116, 2004. \href{https://doi.org/10.1145/989863.989880}
{doi: {{%
10\hspace{.1pt}\discretionary{.}{%
}{.}\hspace{.4pt}1145\discretionary{/}{%
}{/}989863\hspace{.1pt}\discretionary{.}{%
}{.}\hspace{.4pt}989880}}}


\bibitem{saket2016beyond}
B.~Saket, A.~Endert, and J.~Stasko.
\newblock Beyond usability and performance: A review of user experience-focused
  evaluations in visualization.
\newblock In {\em Proc. Workshop on BEyond time and Errors: Novel Evaluation
  Methods for Information Visualization}, pp. 133--142, 2016.
  \href{https://doi.org/10.1145/2993901.2993903}
{doi: {{%
10\hspace{.1pt}\discretionary{.}{%
}{.}\hspace{.4pt}1145\discretionary{/}{%
}{/}2993901\hspace{.1pt}\discretionary{.}{%
}{.}\hspace{.4pt}2993903}}}


\bibitem{scapin1997ergonomic}
D.~L. Scapin and J.~C. Bastien.
\newblock Ergonomic criteria for evaluating the ergonomic quality of
  interactive systems.
\newblock {\em Behaviour \& Information Technology}, 16(4-5):220--231, 1997.
  \href{https://doi.org/10.1080/014492997119806}
{doi: {{%
10\hspace{.1pt}\discretionary{.}{%
}{.}\hspace{.4pt}1080\discretionary{/}{%
}{/}014492997119806}}}


\bibitem{shneiderman2006strategies}
B.~Shneiderman and C.~Plaisant.
\newblock Strategies for evaluating information visualization tools:
  multi-dimensional in-depth long-term case studies.
\newblock In {\em Proceedings of the 2006 AVI workshop on BEyond time and
  errors: novel evaluation methods for information visualization}, pp. 1--7,
  2006. \href{https://doi.org/10.1145/1168149.1168158}
{doi: {{%
10\hspace{.1pt}\discretionary{.}{%
}{.}\hspace{.4pt}1145\discretionary{/}{%
}{/}1168149\hspace{.1pt}\discretionary{.}{%
}{.}\hspace{.4pt}1168158}}}


\bibitem{sperrle2022lotse}
F.~Sperrle, D.~Ceneda, and M.~El-Assady.
\newblock Lotse: A practical framework for guidance in visual analytics.
\newblock {\em IEEE Trans. Visualization and Computer Graphics},
  29(1):1124--1134, 2022. \href{https://doi.org/10.1109/TVCG.2022.3209393}
{doi: {{%
10\hspace{.1pt}\discretionary{.}{%
}{.}\hspace{.4pt}1109\discretionary{/}{%
}{/}TVCG\hspace{.1pt}\discretionary{.}{%
}{.}\hspace{.4pt}2022\hspace{.1pt}\discretionary{.}{%
}{.}\hspace{.4pt}3209393}}}


\bibitem{sperrle2020learning}
F.~Sperrle, A.~V. Jeitler, J.~Bernard, D.~A. Keim, and M.~El-Assady.
\newblock Learning and teaching in co-adaptive guidance for mixed-initiative
  visual analytics.
\newblock In {\em EuroVis Workshop on Visual Analytics (EuroVA)}, pp. 61--65,
  2020. \href{https://doi.org/10.2312/eurova.20201088}
{doi: {{%
10\hspace{.1pt}\discretionary{.}{%
}{.}\hspace{.4pt}2312\discretionary{/}{%
}{/}eurova\hspace{.1pt}\discretionary{.}{%
}{.}\hspace{.4pt}20201088}}}


\bibitem{spinner2019explainer}
T.~Spinner, U.~Schlegel, H.~Sch{\"a}fer, and M.~El-Assady.
\newblock {explAIner}: A visual analytics framework for interactive and
  explainable machine learning.
\newblock {\em IEEE Trans. Visualization and Computer Graphics},
  26(1):1064--1074, 2019. \href{https://doi.org/10.1109/TVCG.2019.2934629}
{doi: {{%
10\hspace{.1pt}\discretionary{.}{%
}{.}\hspace{.4pt}1109\discretionary{/}{%
}{/}TVCG\hspace{.1pt}\discretionary{.}{%
}{.}\hspace{.4pt}2019\hspace{.1pt}\discretionary{.}{%
}{.}\hspace{.4pt}2934629}}}


\bibitem{stasko2014value}
J.~Stasko.
\newblock Value-driven evaluation of visualizations.
\newblock In {\em Proc. Workshop on BEyond time and Errors: Novel Evaluation
  Methods for Information Visualization}, pp. 46--53, 2014.
  \href{https://doi.org/10.1145/2669557.2669579}
{doi: {{%
10\hspace{.1pt}\discretionary{.}{%
}{.}\hspace{.4pt}1145\discretionary{/}{%
}{/}2669557\hspace{.1pt}\discretionary{.}{%
}{.}\hspace{.4pt}2669579}}}


\bibitem{stoiber2022perspectives}
C.~Stoiber, D.~Ceneda, M.~Wagner, V.~Schetinger, T.~Gschwandtner, M.~Streit,
  S.~Miksch, and W.~Aigner.
\newblock Perspectives of visualization onboarding and guidance in {VA}.
\newblock {\em Visual Informatics}, 6(1):68--83, 2022.
  \href{https://doi.org/10.1016/j.visinf.2022.02.005}
{doi: {{%
10\hspace{.1pt}\discretionary{.}{%
}{.}\hspace{.4pt}1016\discretionary{/}{%
}{/}j\hspace{.1pt}\discretionary{.}{%
}{.}\hspace{.4pt}visinf\hspace{.1pt}\discretionary{.}{%
}{.}\hspace{.4pt}2022\hspace{.1pt}\discretionary{.}{%
}{.}\hspace{.4pt}02\hspace{.1pt}\discretionary{.}{%
}{.}\hspace{.4pt}005}}}


\bibitem{wall2018heuristic}
E.~Wall, M.~Agnihotri, L.~Matzen, K.~Divis, M.~Haass, A.~Endert, and J.~Stasko.
\newblock A heuristic approach to value-driven evaluation of visualizations.
\newblock {\em IEEE Trans. Visualization and Computer Graphics},
  25(1):491--500, 2018. \href{https://doi.org/10.1109/TVCG.2018.2865146}
{doi: {{%
10\hspace{.1pt}\discretionary{.}{%
}{.}\hspace{.4pt}1109\discretionary{/}{%
}{/}TVCG\hspace{.1pt}\discretionary{.}{%
}{.}\hspace{.4pt}2018\hspace{.1pt}\discretionary{.}{%
}{.}\hspace{.4pt}2865146}}}


\bibitem{wang2000guidelines}
M.~Q. Wang~Baldonado, A.~Woodruff, and A.~Kuchinsky.
\newblock Guidelines for using multiple views in information visualization.
\newblock In {\em Proc. Int. Conf. on Advanced Visual Interfaces}, pp.
  110--119, 2000. \href{https://doi.org/10.1145/345513.345271}
{doi: {{%
10\hspace{.1pt}\discretionary{.}{%
}{.}\hspace{.4pt}1145\discretionary{/}{%
}{/}345513\hspace{.1pt}\discretionary{.}{%
}{.}\hspace{.4pt}345271}}}


\bibitem{kanit2016voyager}
K.~Wongsuphasawat, D.~Moritz, A.~Anand, J.~Mackinlay, B.~Howe, and J.~Heer.
\newblock Voyager: Exploratory analysis via faceted browsing of visualization
  recommendations.
\newblock {\em IEEE Trans. Visualization and Computer Graphics},
  22(1):649--658, 2015. \href{https://doi.org/10.1109/TVCG.2015.2467191}
{doi: {{%
10\hspace{.1pt}\discretionary{.}{%
}{.}\hspace{.4pt}1109\discretionary{/}{%
}{/}TVCG\hspace{.1pt}\discretionary{.}{%
}{.}\hspace{.4pt}2015\hspace{.1pt}\discretionary{.}{%
}{.}\hspace{.4pt}2467191}}}


\bibitem{wongsuphasawat2017voyager}
K.~Wongsuphasawat, Z.~Qu, D.~Moritz, R.~Chang, F.~Ouk, A.~Anand, J.~Mackinlay,
  B.~Howe, and J.~Heer.
\newblock Voyager 2: Augmenting visual analysis with partial view
  specifications.
\newblock In {\em Proc. SIGCHI Conf. on Human Factors in Computing Systems},
  pp. 2648--2659. ACM, 2017. \href{https://doi.org/10.1145/3025453.3025768}
{doi: {{%
10\hspace{.1pt}\discretionary{.}{%
}{.}\hspace{.4pt}1145\discretionary{/}{%
}{/}3025453\hspace{.1pt}\discretionary{.}{%
}{.}\hspace{.4pt}3025768}}}


\bibitem{zuk2006theoretical}
T.~Zuk and S.~Carpendale.
\newblock Theoretical analysis of uncertainty visualizations.
\newblock In {\em Visualization and Data Analysis}, vol. 6060, pp. 66--79.
  SPIE, 2006. \href{https://doi.org/10.1117/12.643631}
{doi: {{%
10\hspace{.1pt}\discretionary{.}{%
}{.}\hspace{.4pt}1117\discretionary{/}{%
}{/}12\hspace{.1pt}\discretionary{.}{%
}{.}\hspace{.4pt}643631}}}


\bibitem{zuk2006heuristics}
T.~Zuk, L.~Schlesier, P.~Neumann, M.~S. Hancock, and S.~Carpendale.
\newblock Heuristics for information visualization evaluation.
\newblock In {\em Proc. Workshop on BEyond time and Errors: Novel Evaluation
  Methods for Information Visualization}, pp. 1--6, 2006.
  \href{https://doi.org/10.1145/1168149.1168162}
{doi: {{%
10\hspace{.1pt}\discretionary{.}{%
}{.}\hspace{.4pt}1145\discretionary{/}{%
}{/}1168149\hspace{.1pt}\discretionary{.}{%
}{.}\hspace{.4pt}1168162}}}


\end{thebibliography}

\end{document}